\documentclass[a4paper,UKenglish,cleveref]{lipics-v2021}
\hideLIPIcs
\nolinenumbers
\usepackage{mathtools}
\usepackage{xcolor}
\usepackage{etoolbox}

\usepackage[normalem]{ulem}
\usepackage[]{todonotes}

\title{A Non-Dichotomy for CSPs of Stable $\omega$-Categorical Structures}
\titlerunning{A Non-Dichotomy for Stable $\omega$-Categorical CSPs}

\author{Manuel Bodirsky}
  {TU Dresden, Dresden, Germany}
  {manuel.bodirsky@tu-dresden.de}
  {https://orcid.org/0000-0001-8228-3611}{}

\author{David Bradley-Williams}
  {TU Dresden, Dresden, Germany
   \and Czech Technical University in Prague, Prague, Czech Republic
   \and Charles University, Prague, Czech Republic}
  {david.bradley-williams@tu-dresden.de}
  {https://orcid.org/0000-0002-0432-5238}{}

\author{Ehud Hrushovski}
  {University of Oxford, Oxford, United Kingdom}
  {ehud.hrushovski@maths.ox.ac.uk}{}{}

\author{Christoph Spiess}
  {TU Wien, Vienna, Austria}
  {christoph.spiess@tuwien.ac.at}{https://orcid.org/0009-0000-4977-8962}{}

\authorrunning{M. Bodirsky, D. Bradley-Williams, E. Hrushovski, and C. Spiess}

\Copyright{Manuel Bodirsky, David Bradley-Williams, Ehud Hrushovski, and Christoph Spiess}

\funding{
  Funded by the European Union (project POCOCOP, ERC Synergy grant No.~101071674). Views and opinions expressed are however those of the author(s) only and do not necessarily reflect those of the European Union or the European Research Council Executive Agency. Neither the European Union nor the granting authority can be held responsible for them. The research of the second author has also been supported by project 25-15571S of GAČR. The last author’s contribution to this work was funded within
  the DOC program of the OeAW.
}

\acknowledgements{
  The authors thank Bertalan Bodor, David Evans, and Paolo Marimon for discussions on the topic. The core of the proof of Lemma \ref{lem:switch-main} is adapted from an argument that was discovered, in discussion with ChatGPT, during the Bachelor thesis project of Noah Pahlke at the TU Dresden, where the statement of Theorem~\ref{thm:continuum-intro} is also deduced.
}



\newcommand{\NN}{\mathbb{N}}

\newcommand{\QQ}{\mathbb{Q}}
\newcommand{\CSP}{\operatorname{CSP}}
\newcommand{\DTIME}{\operatorname{DTIME}}
\DeclareRobustCommand{\NP}{\ensuremath{\operatorname{NP}}}

\newcommand{\Age}{\operatorname{Age}}
\newcommand{\Aut}{\operatorname{Aut}}
\newcommand{\cl}{\operatorname{cl}}
\newcommand{\ignore}[1]{}
\newcommand{\struct}[1]{\mathfrak{#1}}

\usepackage{pdfcomment}

\newtheorem{question}[theorem]{Question}

\begin{document}

\maketitle 
\begin{abstract}
We construct countable, stable, $\omega$-categorical structures
with at most two binary relations whose CSPs exhibit three
complexity regimes: NP-complete, decidable but not in NP,
and undecidable. The templates arise from a controlled
Hrushovski construction of stable $\omega$-categorical graphs
$\struct{M}_\alpha$ for suitable irrational parameters $\alpha>1$.
Some of these graphs have NP-complete CSPs. Expanding the graphs $\struct{M}_\alpha$ by the non-edge relation turns age membership into a restriction of the CSP.
For every time-constructible bound $t$, there is such an expansion whose CSP is decidable but not in $\DTIME(t)$. For an explicit parameter $\alpha$, the restriction to
complete instances belongs to $\mathrm{coNP}$ and the
full CSP belongs to $\Sigma_2^{\mathrm P}$.
Separating the ages of the $\struct{M}_\alpha$ using graphs with large girth and
prescribed density yields $2^{\aleph_0}$ distinct CSPs
of these expansions, all but countably many of which
are undecidable.
\end{abstract}
\vspace{0.2cm}

\noindent\textbf{AI disclosure.}
OpenAI's ChatGPT was used for proof search and initial LaTeX drafting; it contributed the ideas of using high-girth graphs for the separation argument, applying PCSP hardness results to prove NP-hardness, and using the Kostochka--Yancey bound on the number of edges in critical graphs to establish finite homomorphic equivalents. Many of the other ideas, e.g., the original idea to reconstruct the parameter $\alpha$ from the age, came from the authors, and some emerged in discussions with ChatGPT. The authors reviewed and revised all text produced by LLMs; almost all of the original latex text generated by AI has been changed in the process of revising the document. The authors verified the mathematical content, claims, citations, and final text and take full responsibility for the paper.

\setcounter{tocdepth}{1}
\begingroup\small
\tableofcontents
\endgroup


\section{Introduction}
A constraint satisfaction problem is specified by a fixed relational structure $\struct{B}$ in a finite relational signature, called the \emph{template}.  The input is a finite structure $\struct{A}$ in the same signature, and the question is whether there is a homomorphism $\struct{A}\to\struct{B}$. This problem is denoted by $\CSP(\struct{B})$. All templates considered in this paper have finite relational signature. For finite templates the dichotomy theorem states that each such problem is either in P or NP-complete~\cite{Zhuk20}, announced independently by Bulatov~\cite{Bulatov17} and by Zhuk~\cite{ZhukFVConjecture}. For unrestricted infinite templates, by contrast, every computational decision problem is polynomial-time Turing equivalent to a CSP, so essentially arbitrary computational behaviour can occur \cite[Theorem~1]{BodirskyGrohe08}.  A central aim of infinite-domain CSP theory is therefore to identify structural hypotheses that retain enough finiteness for classification while still covering genuinely infinite domains.

The most established such hypothesis is $\omega$-categoricity.  By the theorem of Ryll-Nardzewski--Engeler--Svenonius (see, e.g.,~\cite[Theorem~7.3.1]{Hod93}), a countable structure is $\omega$-categorical if and only if it has only finitely many automorphism orbits on $n$-tuples for each fixed $n$. Equivalently, its first-order theory has only finitely many $n$-types over the empty set \cite[Theorem~7.3.1]{Hod93}.  Thus, there are only finitely many possible solutions for each input size (up to symmetry), although the domain itself is infinite.
The \emph{universal-algebraic approach} to CSPs, initiated in~\cite{Jeavons,JBK} and further developed in~\cite{wonderland}, played a central role in the proof of the mentioned finite-domain dichotomy theorem. One motivation for focusing on $\omega$-categorical templates is that this approach extends to that setting~\cite{BodirskyNesetrilJLC,Topo-Birk}.\footnote{For example, the computational complexity of $\CSP(\struct{B})$ is captured by the set of \emph{polymorphisms} of $\struct{B}$, up to logspace reductions (see, e.g.,~\cite{Bodirsky21}).}.
However, there are $\omega$-categorical directed graphs such that $\CSP(\struct{B})$ is undecidable~\cite{BodirskyNesetrilJLC}. There are also $\omega$-categorical templates in finite relational signatures whose CSPs have arbitrarily high complexity, even while their polymorphisms satisfy non-trivial identities~\cite[Theorem~1.8]{GJKMP}. Moreover, $\omega$-categorical CSPs occur as complete problems at every level of the polynomial hierarchy~\cite{GuzmanProRydval26}. Assuming $\mathrm P\ne\mathrm{NP}$, there are also $\omega$-categorical directed graphs such that $\CSP(\struct{B})$ is \emph{coNP-intermediate}, i.e., in coNP, but neither in P nor coNP-complete~\cite{BodirskyGrohe08}. The following question is still open.

\begin{question}\label{quest:NP-intermediate}
    Is there an $\omega$-categorical structure $\struct{B}$ with a finite relational signature such that $\CSP(\struct{B})$ is NP-intermediate, i.e., in NP, but neither in P nor NP-complete?
\end{question}

One major research programme narrows the class of $\omega$-categorical templates by focusing on \emph{first-order reducts of finitely bounded homogeneous structures}. A \emph{first-order reduct of $\struct{B}$} is a structure which has the same domain as $\struct{B}$ and all its relations are first-order definable in $\struct{B}$. A structure is \emph{homogeneous} if every isomorphism between finite substructures extends to an automorphism, and \emph{finitely bounded} if its \emph{age}, i.e., the class of all finite structures which embed into it, is characterised by finitely many forbidden finite induced substructures. This class extends finite-domain CSPs and is the setting of the
Bodirsky--Pinsker dichotomy conjecture \cite[Conjecture~1.2]{bodirsky2021projective}. For an overview of this programme and its main methodological challenges, see~\cite{Pinsker22}. The relationship between different algebraic formulations of this conjecture is analysed in~\cite{Topo,BKOPP-equations}. Methods for transferring tractability from finite to infinite domains and for classifying CSPs using canonical polymorphisms and Ramsey theory are developed in~\cite{BodirskyMottet16,MottetPinskerSmooth,BodirskyBodorUIPJournal}. Its scope includes all stable homogeneous structures with finite relational signature studied by Lachlan and Cherlin--Lachlan \cite{Lachlan84,CherlinLachlan86}, as well as all \emph{temporal templates}, i.e., structures first-order definable in $(\mathbb Q;<)$, which include unstable structures, and their CSPs satisfy a complexity dichotomy \cite[Theorem~50]{BodirskyKara10}. 

This, however, is only one way to restrict the full class of $\omega$-categorical structures. A different and complementary question is whether intrinsic model-theoretic tameness conditions, imposed directly on arbitrary $\omega$-categorical templates, force algorithmic regularity. From a model-theoretic perspective, the first candidates are NIP and stability.  NIP excludes the uniform coding of arbitrary finite set systems, whereas stability excludes infinite order patterns; the precise definitions are recalled in \Cref{sec:preliminaries} (stability implies NIP).  Since such patterns are natural mechanisms for encoding computational complexity, it is reasonable to ask the following question.

\begin{question}\label{quest:two}
Do $\omega$-categorical NIP, or even stable, templates have 
decidable CSPs?
\end{question}

If we further strengthen the model-theoretic assumption from stability to \emph{monadic stability}, then this question has a positive answer: Bodor~\cite[Theorem~5.0.1]{BodorDiss} proved a P versus NP-complete dichotomy for CSPs of $\omega$-categorical monadically stable templates. It turns out that monadic stability in combination with $\omega$-categoricity is so strong that such structures are first-order reducts of finitely bounded homogeneous structures~\cite[Theorem~4.9]{Bodor24}, and hence fall into the scope of the mentioned programme. In contrast, stable $\omega$-categorical structures are in general not first-order reducts of finitely bounded homogeneous structures, for example the graphs studied in this paper.
 
The $\omega$-categorical templates used for the earlier undecidable
and coNP-intermediate CSPs are not
NIP~\cite{BodirskyNesetrilJLC,BodirskyGrohe08}\footnote{Their ages
contain the bipartite incidence digraphs
$a_S\to b_i\iff i\in S$, for all
$S\subseteq\{1,\ldots,n\}$ and arbitrarily large $n$.}.
The same holds for the encodings in~\cite{GJKMP}.
One can likewise check that the Boolean QCSP construction
in~\cite[Section~5]{GuzmanProRydval26} has the independence property.

We answer Question~\ref{quest:two} negatively in a strong form. Even stable $\omega$-categorical templates in a finite binary signature can have undecidable CSPs.  Thus, stability---and therefore also NIP---does not by itself provide an alternative to the structural restrictions in the finitely bounded homogeneous programme.

\begin{theorem}\label{thm:continuum-intro}
There are $2^{\aleph_0}$ countable stable $\omega$-categorical graphs with
pairwise distinct ages.
\end{theorem}

The established controlled Hrushovski machinery already supplies stable $\omega$-categorical graph constructions \cite{Hrushovski88,Herwig95,EvansBonn} (the Hrushovski construction should not be confused with the Hrushovski encodings from~\cite{GJKMP,GJKMP-conf}, which produce $\omega$-categorical structures that are not stable).  The new point in \Cref{thm:continuum-intro} is the pairwise separation of their ages. This is yet stronger than pairwise non-isomorphism, since non-isomorphic structures can have the same age. The distinction is crucial for the CSP consequences: the complete instances used below encode precisely age membership and hence also separate the CSPs.

The graphs are obtained by a controlled Hrushovski construction introduced in \cite{Hrushovski88}. For an irrational real parameter $\alpha>1$ we use the predimension
\[
  \delta_\alpha(\struct{A})\coloneqq\alpha|\struct{A}|-e(\struct{A})
\]
on finite graphs, where $|\struct{A}|$ is the number of vertices and $e(\struct{A})$ the number of edges. Using a slowly growing control function $F_\alpha$, we construct a countable graph $\struct{M}_\alpha$ whose age consists exactly of the finite graphs $\struct{A}$ satisfying
\[
  \delta_\alpha(\struct{B})\geq F_\alpha(|\struct{B}|)
  \qquad\text{for every nonempty induced subgraph }\struct{B}\subseteq\struct{A}.
\]
We choose parameters for which $F_\alpha$ tends to infinity, ensuring that the size of a finite set's closure is bounded
in terms of its size, which yields $\omega$-categoricity. To distinguish the ages associated with different parameters,
we construct separating graphs using Erd\H{o}s's theorem that girth and chromatic number can simultaneously be arbitrarily
large \cite{Erdos59}. Although Erd\H{o}s's original proof is probabilistic, subsequent work has provided several constructions, including explicit ones, surveyed in \cite{Nesetril13}.

The CSP consequence concerns a binary definable expansion. For a graph $\struct{M}_\alpha$, let $\struct{B}_\alpha\coloneqq(\struct{M}_\alpha,N)$ be the expansion of $\struct{M}_\alpha$ by the non-edge relation $N$ on distinct pairs.  A complete instance of $\CSP(\struct{B}_\alpha)$ has symmetric, irreflexive relations $E$ and $N$ and assigns exactly one of them to each unordered pair of distinct vertices. Any homomorphism from such an instance is injective and preserves both edges and
non-edges, so complete instances test exactly membership in $\Age(\struct{M}_\alpha)$. Consequently, distinct ages give distinct CSP languages.

\begin{theorem}\label{thm:undecidable-intro}
There are $2^{\aleph_0}$ pairwise distinct CSPs of countable stable $\omega$-categorical structures in a signature consisting of two binary relations.  All but countably many of these CSPs are undecidable.
\end{theorem}

Indeed, there are only countably many decidable languages, because Turing machines have finite descriptions. Thus, the continuum-sized family in \Cref{thm:continuum-intro} immediately yields continuum many undecidable CSPs.

We also give an effective refinement. It shows that the phenomenon is not restricted to noncomputable templates: among computably presented parameters one obtains decidable CSPs of arbitrarily high deterministic time complexity. We encode an instance with $n$ variables by one $n\times n$ matrix for each binary relation, giving bit length $\Theta(n^2)$ for our fixed signatures. Time lower bounds are measured in $n$. $\DTIME(t)$ denotes the class of problems decidable in deterministic time $O(t(n))$, and $t$ is time-constructible if a deterministic machine outputs $t(n)$ on input $1^n$ within $O(t(n))$ steps. A real number $\alpha$ is \emph{computable} if an algorithm, on input $s$, outputs a rational number $q_s$ satisfying $|\alpha-q_s|\leq 2^{-s}$.

\begin{theorem}\label{thm:hierarchy-intro}
For every time-constructible $t\colon\NN\to\NN$ with $t(n)\ge n$, there is a computable real parameter $\alpha >1$ such that $\struct{B}_\alpha$ is stable and $\omega$-categorical, $\CSP(\struct{B}_\alpha)$ is decidable, and
\[
  \CSP(\struct{B}_\alpha)\notin\DTIME(t).
\]
Here the input-size parameter $n$ is the domain size of the instance. The lower bound already holds on complete instances.
\end{theorem}

For parameters admitting polynomial-time comparison with rationals of bounded numerator and denominator, we also obtain a complementary upper bound: the restriction to complete instances belongs to $\mathrm{coNP}$, while the full CSP belongs to $\Sigma_2^{\mathrm P}$. Pantano's construction \cite[Appendix~B]{Pantano1995} provides an explicit such parameter; see \Cref{thm:sigma-two-upper-bound} and \Cref{cor:explicit-upper-bound}.

For the graph reducts, we prove that $\CSP(\struct{M}_\alpha)$ is NP-hard for every $\alpha\in\mathcal A$, and is NP-complete throughout $\mathcal A\cap\bigl((1,5/3)\cup(2,\infty)\bigr)$ (\Cref{thm:graph-hardness} and \Cref{prop:graph-intervals}). Consequently, for all but countably many $\alpha\in\mathcal A\cap \bigl((1,5/3)\cup(2,\infty)\bigr)$, adding the definable non-edge relation $N$ changes an NP-complete CSP into an undecidable one. Hence, stable $\omega$-categorical templates in binary signatures already exhibit three unconditionally distinct algorithmic regimes.

\begin{corollary}[An unconditional non-dichotomy]
\label[corollary]{cor:three-regimes}
There are countably infinite stable $\omega$-categorical structures in signatures with at most two binary relations whose CSPs are, respectively, NP-complete, decidable but not in NP, and undecidable. In the middle case, the time lower bound used to exclude NP is measured by the number of variables. 

Moreover, there is some computable irrational $\alpha$ such that the templates in the first and second case can be chosen to be $\struct{M}_{\alpha}$ and $\struct{B}_{\alpha}$, respectively.
\end{corollary}

\subsection*{Outline}
\Cref{sec:preliminaries} recalls the background. \Cref{sec:predimension} produces stable $\omega$-categorical
graphs by controlled Hrushovski constructions (details
in \Cref{app:generic-construction,app:controlled-details,app:stability}).
\Cref{sec:ages} separates ages uniformly on parameter intervals
using graphs of high girth and prescribed density.
\Cref{sec:csp} links complete instances of non-edge expansions
to age membership and constructs decidable CSPs beyond
prescribed time bounds using nested intervals.
\Cref{sec:colouring-complexity} derives complexity
results for graph reducts from hardness for promise CSPs
and homomorphic equivalence to finite cliques.

\section{Preliminaries}\label{sec:preliminaries}
Our construction uses embeddings to control the ages of
the templates, while their CSPs use homomorphisms.
We first fix these notions, then recall the tameness
conditions and the amalgamation framework used below.

We use $\NN=\{0,1,2,\ldots\}$ and write $x_n\nearrow x$
for non-decreasing convergence.

\paragraph*{Structures and maps.}
We work in finite relational signatures.
We write $A$ for the domain of $\struct{A}$ and
$|\struct{A}|$ for its cardinality. For structures $\struct{A}$ and $\struct{B}$ in the same signature $\tau$, we call $\struct{A}$ a \emph{substructure} of $\struct{B}$, written $\struct{A}\subseteq\struct{B}$, if $A\subseteq B$ and $R^{\struct{A}}=R^{\struct{B}}|_{A^k}$ for every $k$-ary relation $R\in\tau$. Thus, substructures are induced. We identify domain subsets with the induced substructures they determine. For substructures of a common structure, unions, intersections, and differences are taken on their domains and equipped with the induced relations.

For two structures $\struct{A}$ and $\struct{B}$ in the same signature $\tau$, a homomorphism from $\struct{A}$ to $\struct{B}$ is a map $h\colon A\to B$ that preserves every relation: for every $k$-ary $R\in\tau$ and every $(a_1,\ldots,a_k)\in A^k$, we have $(a_1,\ldots,a_k)\in R^{\struct{A}} \Longrightarrow (h(a_1),\ldots,h(a_k))\in R^{\struct{B}}.$
We write $\struct{A}\to\struct{B}$ if there exists a homomorphism from $\struct{A}$ to $\struct{B}$. An embedding is an injective homomorphism that additionally satisfies $(h(a_1),\ldots,h(a_k))\in R^{\struct{B}} \Longrightarrow (a_1,\ldots,a_k)\in R^{\struct{A}}$. An isomorphism is a surjective embedding.
An automorphism is an isomorphism from a structure to itself; $\Aut(\struct{M})$ denotes its automorphism group.
The problem $\CSP(\struct{B})$ asks whether a finite input structure in the same signature admits a homomorphism to $\struct{B}$ \cite[Definition~1.1.1]{Bodirsky21}.

For two structures $\struct{A}$ and $\struct{B}$
on the same domain with signatures $\tau_1$ and $\tau_2$,
respectively, we say $\struct{B}$ is an \emph{expansion}
of $\struct{A}$ and, conversely, $\struct{A}$ is a
\emph{reduct} of $\struct{B}$, if $\tau_1\subseteq\tau_2$
and $R^{\struct{A}}=R^{\struct{B}}$ for every $R\in\tau_1$.
An expansion is \emph{first-order definable} if every
added relation is first-order definable without
parameters in the reduct. 
The \emph{age} of a structure $\struct{M}$, denoted 
$\Age(\struct{M})$, consists of the finite induced
substructures of $\struct{M}$, up to isomorphism.

Unless explicitly called a \emph{directed graph} (or \emph{digraph}),
a graph is simple and undirected, represented by a symmetric,
irreflexive binary relation $E$.
CSP input structures may have loops and asymmetric binary relations.
For our graph templates and their non-edge expansions, a loop in any
input relation makes the instance unsatisfiable, while adding the
converse of every constraint leaves its homomorphisms unchanged.
For a graph $\struct{H}$, let
$V(\struct{H})$ be its vertex set and $E(\struct{H})$
its set of unordered edges.
We write $|\struct{H}|=|V(\struct{H})|$ and
$e(\struct{H})=|E(\struct{H})|$.
The term \emph{subgraph} also allows edge deletions;
we specify when such a subgraph is induced. The \emph{density} of $\struct{H}$ is $e(\struct{H})/|\struct{H}|$. We write $\chi(\struct{H})$ for the chromatic number of $\struct{H}$.

\paragraph*{Model-theoretic tameness.}
A countable structure $\struct{M}$ is
\emph{$\omega$-categorical} if its theory has exactly
one countable model up to isomorphism.
By Ryll-Nardzewski--Engeler--Svenonius, this is
equivalent to $\Aut(\struct{M})$ having finitely many
orbits on $M^n$ for every $n$
\cite[Theorem~7.3.1]{Hod93}, where two tuples lie in the same automorphism orbit if an automorphism maps one to the other componentwise.

A theory is \emph{stable} if no formula
$\varphi(\bar x;\bar y)$ has the \emph{order property}:
there are tuples $(\bar a_i)_{i\in\NN}$ and
$(\bar b_j)_{j\in\NN}$ in a model of the theory such
that $\varphi(\bar a_i;\bar b_j)$ holds exactly when
$i<j$.
A formula has the \emph{independence property} if,
for every $n$, there are tuples
$\bar b_1,\ldots,\bar b_n$ and $\bar a_S$, for all
$S\subseteq\{1,\ldots,n\}$, in a model of the theory,
such that $\varphi(\bar a_S;\bar b_i)$ holds exactly when $i\in S$.
A theory is \emph{NIP} if no formula has this property. It is well-known and easy to see that stability implies NIP~(see Tent-Ziegler~\cite[Exercise 8.2.2]{TZ12}). We say that a structure $\struct{M}$ is stable (respectively, NIP) if its first-order theory $\operatorname{Th}(\struct{M})$ is stable (respectively, NIP).

\begin{remark}\label[remark]{rem:definable-expansions}
Expansions by relations that are first-order definable without
parameters preserve both $\omega$-categoricity and
stability. Such an expansion has the same automorphism
group, so $\omega$-categoricity is preserved by
Ryll-Nardzewski--Engeler--Svenonius.
Every formula in the expansion translates to a formula
in the reduct, so the expansion cannot introduce the
order property.
\end{remark}

\subsection{Constructing countable structures}
We briefly recall the variant of Fra\"{\i}ss\'{e} amalgamation that we employ for constructing the countable CSP templates, following the presentation of Evans \cite[Section~3]{EvansBonn}, making use of the variant of Fra\"{\i}ss\'{e}'s theorem given in \cite[Theorem~2.10]{Evans94}, similar to that stated in \cite[Theorem~1.5]{KuekerLaskowski1992}.

Let $\mathcal K$ be a class of finite structures in a fixed finite relational signature. The class is \emph{hereditary} if every substructure of a member of $\mathcal K$ also belongs to $\mathcal K$. Let $\leq$ be a reflexive and transitive relation on $\mathcal K$, preserved by isomorphisms, such that $\struct{A}\leq\struct{B}$ implies $\struct{A}\subseteq\struct{B}$. If we choose $\leq$ to be the (induced) substructure relation, then we recover the classical theory of Fra\"{\i}ss\'{e} amalgamation. We call $\struct{A}$ a \emph{strong substructure} of $\struct{B}$ if $\struct{A}\leq\struct{B}$. We assume the following \emph{convexity condition}:
\[
  \struct{A}\leq\struct{B},\qquad \struct{A}\subseteq\struct{C}\subseteq\struct{B},\qquad \struct{C}\in\mathcal K
  \quad\Longrightarrow\quad \struct{A}\leq\struct{C}.
\]
This is condition~(N3) in \cite{EvansBonn}.

An embedding $f\colon\struct{A}\to\struct{B}$ is \emph{strong} if $f(\struct{A})\leq\struct{B}$. The \emph{joint embedding property} means that any two members of $\mathcal K$ strongly embed into a common member. The \emph{amalgamation property} means that, for strong embeddings $f_i\colon\struct{A}\to\struct{B}_i$ between members of $\mathcal K$, there are $\struct{C}\in\mathcal K$ and strong embeddings $g_i\colon\struct{B}_i\to\struct{C}$ such that $g_1\circ f_1=g_2\circ f_2$; the resulting structure and maps form an \emph{amalgam}.

We call $(\mathcal K,\leq)$ an \emph{amalgamation class} if $\mathcal K$ is hereditary and closed under isomorphisms, has countably many isomorphism types, and has the joint embedding and amalgamation properties. If $\Age(\struct{M})\subseteq\mathcal K$, a finite substructure $\struct{A}\subseteq\struct{M}$ is \emph{strong in $\struct{M}$}, written $\struct{A}\leq\struct{M}$, if $\struct{A}\leq\struct{B}$ for every finite substructure $\struct{B}\subseteq\struct{M}$ containing $\struct{A}$. An embedding of a finite structure into $\struct{M}$ is \emph{strong} if its image is strong in $\struct{M}$.

Convexity and transitivity imply that
\[
  \struct{A}\leq\struct{B}\leq\struct{M} \quad\Longrightarrow\quad \struct{A}\leq\struct{M}
\]
whenever $\struct{A}$ and $\struct{B}$ are finite. Indeed, given a finite substructure $\struct{D}\subseteq\struct{M}$ containing $\struct{A}$, let $\struct{E}\coloneqq\struct{B}\cup\struct{D}$. Then $\struct{B}\leq\struct{E}$ since $\struct{B}$ is strong in $\struct{M}$, so $\struct{A}\leq\struct{E}$ by transitivity, and hence $\struct{A}\leq\struct{D}$ by convexity.

\begin{theorem}[{Amalgamation limit; see \cite[Theorem~2.10]{Evans94} and \cite[Theorem~3.2]{EvansBonn}}]
\label[theorem]{thm:am_lim}
Let $(\mathcal K,\leq)$ be an amalgamation class containing structures of arbitrarily large finite size. Let $\leq$ be reflexive, transitive, and satisfy the convexity condition. Then there exists an, up to isomorphism, unique countably infinite structure $\struct{M}$, called the \emph{amalgamation limit} of $(\mathcal K,\leq)$, satisfying:
\begin{enumerate}[(i)]
\item \label{ageamalgamationlimit} $\Age(\struct{M})=\mathcal K;$
\item \label{finitesubsetcontainedinstrongsubstr} Every finite subset of $\struct{M}$ is contained in a finite strong substructure of $\struct{M}$;
\item \textup{(Extension property)}\label{extensionamalgamationlimit} For all $\struct{A},\struct{B}\in\mathcal K$ with $\struct{A}\leq\struct{B}$, every strong embedding $f\colon\struct{A}\to\struct{M}$ extends to a strong embedding $\struct{B}\to\struct{M}$.
\end{enumerate}
Moreover, $\struct{M}$ satisfies the following weakening of homogeneity: every isomorphism between finite strong substructures of $\struct{M}$ extends to an automorphism of $\struct{M}$.
\end{theorem}

Amalgamation limits are variously called \emph{generic structures} or \emph{Fra{\"i}ss{\'e} limits} in the literature. The theorem here is the version of Fra\"{\i}ss\'{e} amalgamation with strong embeddings presented in \cite[Theorem~2.10]{Evans94} and \cite[Theorem~3.2]{EvansBonn}. We recall its proof in \Cref{app:generic-construction}. These results can also be viewed as instances of more general categorical versions of Fra\"{\i}ss\'{e}'s theorem, where limits arise as colimits of chains of finite structures, with strong embeddings as morphisms; see, e.g.,
\cite{Caramello2014,DrosteGoebel1992,DrosteGoebel1993,Kubis2014,Kubis2022}. All of the above generalise the theorem of Fra\"{\i}ss\'{e} \cite{Fraisse1953,Fraisse1954}; see, e.g.,~\cite[Chapter~11]{Fraisse1986} and \cite[Theorem~7.1.2]{Hod93}.

Homogeneity with respect to strong substructures yields $\omega$-categoricity provided every finite tuple lies in a strong substructure whose size is bounded in terms of the tuple length. We shall use the following criterion.

\begin{lemma}[Bounded strong substructures]
\label[lemma]{lem:bounded-strong-substructures}
Let $\struct{M}$ be the amalgamation limit of $(\mathcal K,\leq)$ in a finite relational signature. Suppose there is a function $b\colon\mathbb N\to\mathbb N$ such that, for every $\struct{B}\in\mathcal K$, each subset $\struct{A}\subseteq\struct{B}$ is contained in a strong substructure $\struct{C}\leq\struct{B}$ with $|\struct{C}|\leq b(|\struct{A}|)$. Then $\struct{M}$ is $\omega$-categorical.
\end{lemma}

\section{Controlled predimension classes}\label{sec:predimension}

This section specialises the established framework of Hrushovski constructions with the graph predimension and parameter-dependent control functions used below. 
We include the details for completeness and convenience of the reader.  The new ingredients begin with the uniform separation of the resulting ages in the next section and its CSP applications. Closely related constructions appear in \cite[Section~3, Section~5, and Section~5.1, respectively] {Herwig95,Wagner94,Pantano1995}. All these constructions are based on Hrushovski's in \cite{Hrushovski88}. We follow the presentation given by Evans~\cite[Section~3]{EvansBonn}. 
Proofs of the standard approximation, strongness, amalgamation, and closure
calculations in this section are collected in
\Cref{app:controlled-details}.

For a real parameter $\alpha>1$ and $N\ge1$, define the distance from
$\alpha$ to its best rational approximation from below with denominator at
most $N$ by
\begin{align}
  \eta(\alpha,N)\coloneqq\min\left\{\alpha-\frac ab:
  a,b\in\NN,\ 1\le b\le N,\ \frac ab<\alpha\right\}.
  \label{eq:eta}
\end{align}
Put
\begin{align}
  F_\alpha(0)\coloneqq0,
  \qquad
F_\alpha(N)\coloneqq\sum_{k=1}^N\eta(\alpha,k)
  \quad\text{for all }N\ge1,
  \label{eq:Falpha}
\end{align}
and extend $F_\alpha$ linearly between consecutive integers.  We call
$F_\alpha$ the \emph{control function}.  The sequence $\eta(\alpha,N)$ is
positive and non-increasing.  The elementary properties needed below are
collected next and proven in \Cref{app:controlled-details}.

\begin{lemma}[Properties of $F_\alpha$]\label[lemma]{lem:approximation-estimates}
For every real $\alpha>1$ and $N\ge1$:
\begin{enumerate}[(i)]
\item $0<\eta(\alpha,N+1)\leq\eta(\alpha,N)\leq 1/N$;
\item $F_\alpha$ is continuous, strictly increasing, and concave;
\item $F_\alpha(N)\leq \sum_{k=1}^N\frac1k$, and hence $F_\alpha(N)/N\to0$.
\end{enumerate}
\end{lemma}

Since $F_\alpha$ is increasing, it is unbounded exactly when $F_\alpha(N)\to\infty$; we use the latter formulation below. 
Given a real $\alpha > 1$, define the \emph{predimension} of a finite graph $\struct{A}$ by
\[
\delta_\alpha(\struct{A})\coloneqq\alpha|\struct{A}|-e(\struct{A}).
\]
For finite induced subgraphs $\struct{B},\struct{C}$ of a common graph, direct edge counting gives the \emph{submodularity inequality}
\begin{equation}\label{eq:submod}
  \delta_\alpha(\struct{B}\cup\struct{C})
  \le \delta_\alpha(\struct{B})+\delta_\alpha(\struct{C})-\delta_\alpha(\struct{B}\cap\struct{C}),
\end{equation}
with equality when there are no edges between $\struct{B}\setminus\struct{C}$ and $\struct{C}\setminus\struct{B}$; see 
\cite[Lemma~3.7]{EvansBonn}.

\begin{definition}\label{def:classC}
For any fixed real $\alpha > 1$, let $\mathcal C_\alpha$ be the class of finite graphs $\struct{A}$ such that, for every non-empty induced $\struct{B}\subseteq\struct{A}$,
\begin{equation}\label{eq:control}
  \delta_\alpha(\struct{B})\ge F_\alpha(|\struct{B}|).
\end{equation}

Given $\struct{A}\subseteq\struct{B} \in \mathcal C_\alpha$, write $\struct{A}\leq_\alpha\struct{B}$ if
\begin{equation}\label{eq:strong}
  \delta_\alpha(\struct{A})<\delta_\alpha(\struct{B}')
  \qquad\text{whenever }\struct{A}\subsetneq\struct{B}'\subseteq\struct{B}.
\end{equation}
In this case we say that $\struct{A}$ is \emph{closed in $\struct{B}$}.
\end{definition}

Deleting edges preserves membership in $\mathcal C_\alpha$: on every vertex subset, it can only increase the predimension, while leaving the right-hand side of \eqref{eq:control} unchanged. Thus, \eqref{eq:control} also holds for subgraphs that are not induced.

 The proof of the following is given in \Cref{app:controlled-details}.

\begin{lemma}[Properties of $\le_\alpha$]\label[lemma]{lem:strongness-calculus}
The relation $\leq_\alpha$ is reflexive and transitive. Moreover, if $\struct{A}\leq_\alpha\struct{B}$ and $\struct{X}\subseteq\struct{B}$, then $\struct{A}\cap\struct{X}\leq_\alpha\struct{X}$.
\end{lemma}

Consequently, finite intersections of subgraphs closed in $\struct{B}$ are closed in $\struct{B}$, and $\struct{B}$ is closed in itself. For finite $\struct{A}\subseteq\struct{B}\in\mathcal C_\alpha$ let $\cl_{\struct{B}}(\struct{A})$ denote the unique smallest closed subgraph of $\struct{B}$ containing $\struct{A}$.

\subsection{Free amalgamation}
The \emph{free amalgam} of $\struct{B}_1$ and $\struct{B}_2$ over their common induced subgraph $\struct{A}$ has vertex set $V(\struct{B}_1)\cup V(\struct{B}_2)$ and edge set $E(\struct{B}_1)\cup E(\struct{B}_2)$. Thus, there are no edges between the two parts outside $\struct{A}$. A \emph{free-amalgamation class} is an amalgamation class in which this free amalgam is always an amalgam for strong embeddings. The proof of the following is given in \Cref{app:controlled-details}.

\begin{lemma}\label[lemma]{lem:amalgamation}
For every irrational $\alpha>1$, the pair $(\mathcal C_\alpha,\leq_\alpha)$ is a free-amalgamation class.
\end{lemma}

In the following, we denote by $\struct{M}_\alpha$ the amalgamation limit of $(\mathcal C_\alpha,\leq_\alpha)$ in the sense of \Cref{thm:am_lim}.

\subsection{Bounded closures and stability}

The following lemma bounds the predimension of the closure defined after \Cref{lem:strongness-calculus}. Its proof is given in \Cref{app:controlled-details}.

\begin{lemma}[Predimension of closure]\label[lemma]{lem:closure-predimension}
$\delta_\alpha(\cl_{\struct{B}}(\struct{A}))\leq\delta_\alpha(\struct{A})$ for every finite $\struct{A}\subseteq\struct{B}\in\mathcal C_\alpha$.
\end{lemma}

Suppose that $F_\alpha(N)\to\infty$, so that $F_\alpha^{-1}$ is defined on $[0,\infty)$. For nonempty $\struct{A}$, \Cref{lem:closure-predimension} and the control inequality give $F_\alpha(|\cl_{\struct{B}}(\struct{A})|)\leq\alpha|\struct{A}|$; for $\struct{A}=\varnothing$, the closure is empty. Consequently,
\begin{equation}\label{eq:closure-bound}
  |\cl_{\struct{B}}(\struct{A})|\le
  \left\lfloor F_\alpha^{-1}(\alpha|\struct{A}|)\right\rfloor.
\end{equation}
This is a uniform bound depending only on $|\struct{A}|$. By \Cref{lem:bounded-strong-substructures}, the amalgamation limit $\struct{M}_\alpha$ is therefore $\omega$-categorical.

Free amalgamation with respect to strong embeddings does not itself imply stability. However, when $F_\alpha(N)\to\infty$ stability of these structures is well established; a proof can be found in \cite{Herwig95} (see the first paragraph of Section 6 with the proof of stability in Section 4 culminating in Theorem 14), and also follows from Wagner's dimension-theoretic analysis \cite[Section~3 and Example~5.3]{Wagner94}. For completeness, we give an account of the dimension calculus and type-counting argument in \Cref{app:stability}, following Wagner's dimension-theoretic approach to stability for predimension generics.

\begin{proposition}\label[proposition]{prop:tameness}
Suppose that $\alpha>1$ is irrational and $F_\alpha(N)\to\infty$. Then $\struct{M}_\alpha$ is $\omega$-categorical and stable.
\end{proposition}

\begin{proof}
The preceding closure argument gives $\omega$-categoricity. Stability is proved in \Cref{thm:appendix-stability}.
\end{proof}

\section{A continuum of distinct ages}\label{sec:ages}
It remains to determine continuum many real parameters $\alpha$ for which $F_\alpha(N)\to\infty$ and to show that the corresponding graphs $\struct{M}_\alpha$ have pairwise distinct ages.

\subsection{Many admissible parameters}\label{sect:admiss}
Evans sketches a Baire-category argument showing that parameters $\alpha$ for which $F_\alpha(N)$ diverges form a comeager set \cite[Example~3.22]{EvansBonn}, also mentioned by Wagner \cite[Example~5.3]{Wagner94}, while Pantano in \cite[Appendix~A]{Pantano1995} gives such details for a related controlled predimension construction in a countably infinite language. For the uniform and effective interval arguments in our complexity analysis, we need the following explicit strengthening; a full proof suited to our conventions is given in \Cref{app:evans-density}.

\begin{lemma}[Uniform density consequence]
\label[lemma]{lem:evans-uniform-density}
For every nonempty open interval $I\subseteq(1,\infty)$ and every $m\in\NN$, there are $K\in\NN$ and a nonempty rational open interval $J$ with $\overline J\subseteq I$ such that
\[
  \sum_{N=1}^K\eta(\alpha,N)>m
  \qquad\text{for every }\alpha\in J.
\]
\end{lemma}

For $m\in\NN$, let $U_m$ be the union of all rational open intervals $J$
for which there is an integer $K$ such that
\begin{equation}\label{eq:dense-uniform}
  \sum_{N=1}^K\eta(\alpha,N)>m
  \qquad\text{for all }\alpha\in J.
\end{equation}
By \Cref{lem:evans-uniform-density}, every $U_m$ is open and dense.

We shall use the following explicit strengthening of the usual Baire
conclusion; its standard interval-tree proof is recalled in
\Cref{app:evans-density}.

\begin{lemma}[Intersection of dense open sets]\label[lemma]{lem:comeager-continuum}
If $V_0,V_1,\ldots$ are dense open subsets of a non-empty open real interval $I$,
then $\bigcap_mV_m$ contains a subset homeomorphic to the Cantor space $2^\omega$, the space of infinite binary sequences with the product topology, where $\{0,1\}$ carries the discrete topology. In particular, the intersection has cardinality $2^{\aleph_0}$.
\end{lemma}

By \Cref{lem:comeager-continuum}, the set of \emph{admissible parameters}
\begin{align}
  \mathcal A\coloneqq((1,\infty)\setminus\QQ)\cap\bigcap_{m\in\NN}U_m
  \label{eq:admissible-parameters}
\end{align}
has cardinality $2^{\aleph_0}$; here the irrationals are themselves the intersection of countably many open dense sets.  For every $\alpha\in\mathcal A$, the partial sums exceed every integer, and hence $F_\alpha(N)\to\infty$. Therefore, \Cref{lem:amalgamation} and \Cref{prop:tameness} apply.

\subsection{The switch lemma}

The \emph{girth} of a graph is the length of its shortest cycle, with girth
$\infty$ for a forest. Erd\H{o}s proved that graphs can have simultaneously arbitrarily large girth and chromatic number \cite{Erdos59}. That was proven with a probabilistic construction. Subsequent work has produced such graphs by a variety of methods, including explicit constructions; see \cite{Nesetril13} for a survey. We use the following density consequence; the argument is given in
\Cref{app:high-girth-density}.

\begin{lemma}[High girth and density]\label[lemma]{lem:high-girth-density}
For every integer $g\ge3$, every real $q>1$, and every $L\in\NN$, there is
a finite graph $\struct{H}$ with at least $L$ vertices, girth greater than $g$, and
$e(\struct{H})>q|\struct{H}|$.
\end{lemma}

The graph obtained in the following lemma is the central separating device.

\begin{lemma}[Switch lemma]\label[lemma]{lem:switch-main}
Let $u_0,u_1,v_0,v_1$ be real numbers such that $1<u_0<u_1<v_0<v_1$. There is a finite graph $\struct{G}$ such that
\begin{enumerate}[(i)]
\item $\struct{G}\notin\mathcal C_\alpha$ for every $\alpha\in(u_0,u_1)\cap\mathcal A$;
\item $\struct{G}\in\mathcal C_\alpha$ for every $\alpha\in(v_0,v_1)\cap\mathcal A$.
\end{enumerate}
A graph satisfying (i) and (ii) is called a \emph{switch graph} for $(u_0,u_1,v_0,v_1)$. It may be chosen with $e(\struct{G})>u_1|\struct{G}|$, arbitrarily large girth, and arbitrarily many vertices.
\end{lemma}

\begin{proof}
Let $q\coloneqq(u_1+v_0)/2$, and set $\gamma\coloneqq\frac{v_0-u_1}{2}=q-u_1=v_0-q$. Choose $g$ so large that
\[
  \frac1g\le\frac\gamma2
  \quad\text{and}\quad
  \sum_{k=1}^m\frac1k\le\frac\gamma2m\quad\text{for all }m\ge g.
\]

By \Cref{lem:high-girth-density}, there are arbitrarily
large graphs $\struct{H}$ of girth greater than $g$
with $e(\struct{H})>q|\struct{H}|$.
Choose a subgraph $\struct{G}$ of $\struct{H}$, not necessarily induced, satisfying $e(\struct{G})>q|\struct{G}|$, such that no proper
subgraph obtained by deleting vertices, edges, or both
satisfies this inequality.
Every proper subgraph $\struct{B}$ of $\struct{G}$
then satisfies $e(\struct{B})\le q|\struct{B}|$, and
\[
  q|\struct{G}|<e(\struct{G})\le q|\struct{G}|+1.
\]

Moreover, as $q>1$, we have $e(\struct{G})>q|\struct{G}|>|\struct{G}|-1$, so $\struct{G}$ contains a cycle. Its girth is greater than $g$, and therefore $|\struct{G}|>g$.

For $\alpha\in(u_0,u_1)$, we have $\alpha<q<e(\struct{G})/|\struct{G}|$, so $\delta_\alpha(\struct{G})<0\le F_\alpha(|\struct{G}|)$, hence, $\struct{G}\notin\mathcal C_\alpha$.

Now let $\alpha\in(v_0,v_1)$ and take any nonempty induced subgraph $\struct{S}\subseteq\struct{G}$. Put $m:=|\struct{S}|$.
If $\struct{S}$ contains a cycle, then $m>g$ and
minimality gives $e(\struct{S})\le qm+1$, including
the case $\struct{S}=\struct{G}$.
Since $\alpha-q>\gamma$, the choice of $g$ gives
\[
  \delta_\alpha(\struct{S})-F_\alpha(m)
  \ge(\alpha-q)m-1-\sum_{k=1}^m\frac1k
  >\frac{\gamma m}{2}-1\ge 0.
\]
If $\struct{S}$ is acyclic, then
$e(\struct{S})\le m-1$. The rational approximation 
$1/1<\alpha$ gives $\eta(\alpha,N)\le\alpha-1$ for
every $N\ge1$, so $F_\alpha(m)\le(\alpha-1)m$ and
\[
  \delta_\alpha(\struct{S})-F_\alpha(m)
  \ge\alpha m-(m-1)-(\alpha-1)m=1.
\]
Thus $\struct{G}\in\mathcal C_\alpha$.
Increasing the initial choice of $g$ gives switch
graphs with girth and number of vertices larger
than any prescribed bounds.
\end{proof}

A switch graph for rational inputs $u_0,u_1,v_0,v_1$ can be found effectively. Take $q\coloneqq(u_1+v_0)/2$ and an integer $g\ge3$ satisfying $1/g\le\gamma/2$ and $(\sum_{k=1}^g\frac1k)/g\le\gamma/2$. Since $(\sum_{k=1}^m\frac1k)/m$ decreases, the required harmonic estimate holds for all $m\ge g$. Enumerate finite graphs in increasing order of their
number of vertices and, for each fixed number of vertices, their number of edges, until finding one with girth greater than $g$ and density greater than $q$. Every proper subgraph would have appeared earlier, up to isomorphism, if it satisfied the same density condition. Thus every proper subgraph $\struct{S}$ of the resulting graph $\struct{G}$ satisfies
$e(\struct{S})\le q|\struct{S}|$. \Cref{lem:high-girth-density} guarantees termination. Alternatively, it may result in a faster implementation to use an explicit construction of graphs with high chromatic number and high girth graphs, c.f.~\cite{Nesetril13}. 

The following proposition shows that one can recover the parameter $\alpha$ from the age of $\struct{M}_\alpha$.

\begin{proposition}[Recovering the parameter]\label[proposition]{prop:recover-parameter}
For every $\alpha\in\mathcal A$,
\[
  \sup\left\{\frac{e(\struct{G})}{|\struct{G}|}:\varnothing\neq\struct{G}\in\mathcal C_\alpha\right\}
  =\alpha.
\]
\end{proposition}

\begin{proof}
If $\struct{G}\in\mathcal C_\alpha$, then $\alpha|\struct{G}|-e(\struct{G})=\delta_\alpha(\struct{G})\geq F_\alpha(|\struct{G}|)\geq0$, so the supremum is at most $\alpha$. Conversely, fix a rational number $q$ with $1<q<\alpha$. Choose rationals $1<q<u_0<u_1<v_0<\alpha<v_1$. A switch graph $\struct{G}$ for $(u_0,u_1,v_0,v_1)$ belongs to $\mathcal C_\alpha$ and satisfies $e(\struct{G})>u_1|\struct{G}|>q|\struct{G}|$. Letting $q\nearrow\alpha$ proves the reverse inequality.
\end{proof}

\begin{proof}[Proof of \Cref{thm:continuum-intro}]
For every $\alpha\in\mathcal A$, the graph $\struct{M}_\alpha$ is stable and $\omega$-categorical by \Cref{prop:tameness}. If $\alpha\neq\beta$, then \Cref{prop:recover-parameter} gives $\mathcal C_\alpha\neq\mathcal C_\beta$, so the ages are pairwise distinct. Since $|\mathcal A|=2^{\aleph_0}$, the result follows.
\end{proof}

\section{Constraint satisfaction consequences}\label{sec:csp}

For $\alpha\in\mathcal A$, expand $\struct{M}_\alpha$ by the relation
\[
  N\coloneqq\{(x,y)\in\struct{M}_\alpha^2:
  x\neq y\text{ and }\neg E(x,y)\},
\]
and write $\struct{B}_\alpha\coloneqq(\struct{M}_\alpha,N)$. This is a structure in a finite binary signature. Since $N$ is first-order definable in $\struct{M}_\alpha$,
the expansion is again stable and $\omega$-categorical by \Cref{rem:definable-expansions}.

Given a finite graph $\struct{G}$, let $\struct{X}_{\struct{G}}$
be the $\{E,N\}$-structure on $V(\struct{G})$ defined by
$E(x,y)\iff\{x,y\}\in E(\struct{G})$ and
$N(x,y)\iff x\neq y\land\neg E(x,y)$.
We call such instances \emph{complete}.

\begin{lemma}[Complete instances detect age membership]
\label[lemma]{lem:complete-age}
For every finite graph $\struct{G}$ and every $\alpha\in\mathcal A$,
\begin{equation}\label{eq:complete-age}
  \struct{X}_{\struct{G}}\in\CSP(\struct{B}_\alpha)
  \quad\Longleftrightarrow\quad
  \struct{G}\in\Age(\struct{M}_\alpha)=\mathcal C_\alpha.
\end{equation}
\end{lemma}
Irreflexivity forces injectivity, and preserving both relations gives an induced embedding; see \Cref{app:effective-details} for details. Combining this observation with \Cref{prop:recover-parameter} shows that the CSP language itself remembers the parameter already on complete inputs.

\begin{corollary}[Recovering the parameter from the CSP]
\label[corollary]{cor:csp-recovers-parameter}
For every $\alpha\in\mathcal A$,
\[
  \alpha=
  \sup\left\{\frac{e(\struct{G})}{|\struct{G}|}:\struct{G}\neq\varnothing,
  \ \struct{X}_{\struct{G}}\in\CSP(\struct{B}_\alpha)\right\}.
\]
In particular, $\alpha\neq\beta$ implies $\CSP(\struct{B}_\alpha)\neq\CSP(\struct{B}_\beta)$.
\end{corollary}

\begin{proof}
By \Cref{lem:complete-age}, the complete inputs accepted by the CSP are precisely the structures $\struct{X}_{\struct{G}}$ with  $\struct{G}\in\mathcal C_\alpha$. The equality is therefore \Cref{prop:recover-parameter}. Distinct values of the displayed supremum force distinct languages.
\end{proof}

Having obtained pairwise distinctness of the CSPs, we are now ready to prove \Cref{thm:undecidable-intro}. 

\begin{proof}[Proof of \Cref{thm:undecidable-intro}]
By \Cref{thm:continuum-intro}, the family $(\struct{M}_\alpha)_{\alpha\in\mathcal A}$ contains $2^{\aleph_0}$ countable stable $\omega$-categorical graphs with pairwise distinct ages. Since $N$ is first-order definable from $E$, \Cref{rem:definable-expansions} shows that every $\struct{B}_\alpha=(\struct{M}_\alpha,N)$ is again stable and $\omega$-categorical. \Cref{cor:csp-recovers-parameter} shows that the corresponding CSPs are pairwise different. Hence the family contains $2^{\aleph_0}$ distinct languages. Only countably many languages are decided by Turing machines, so all but countably many members of the family are undecidable.
\end{proof}

For arbitrary instances, homomorphisms may identify unconstrained variables. The following finite characterisation isolates this extra freedom; its proof is given in \Cref{app:effective-details}.

\begin{lemma}[Quotients]\label[lemma]{lem:quotient-completion}
Let $\struct{X}$ be an instance of $\CSP(\struct{B}_\alpha)$. For an equivalence relation $\sim$ on $X$, let the quotient graph $\struct{G}_\sim$ be the simple undirected graph whose vertices are the $\sim$-classes and whose edges join exactly the distinct classes linked by an $E$-constraint in either direction in $\struct{X}$. Then $\struct{X}\to\struct{B}_\alpha$ if and only if there is such an equivalence relation $\sim$ satisfying:
\begin{enumerate}[(i)]
\item no ordered pair in $E^{\struct{X}}$ or in $N^{\struct{X}}$ is contained in $\sim$;
\item there are no $x_1\sim x_2$ and $y_1\sim y_2$ such that $E(x_1,y_1)$ and either $N(x_2,y_2)$ or $N(y_2,x_2)$;
\item $\struct{G}_\sim\in\mathcal C_\alpha$.
\end{enumerate}
We call $\sim$ \emph{consistent} if it satisfies
\textup{(i)} and \textup{(ii)}.
\end{lemma}

For $\alpha>1$, $j\ge1$, and $n\in\NN$, write
\begin{equation}
\label{eq:lower-approximations}
  r_\alpha(j):=\alpha-\eta(\alpha,j),
  \qquad
  S_\alpha(n):=\sum_{j=1}^{n}r_\alpha(j)
             =\alpha n-F_\alpha(n).
\end{equation}
Thus, $r_\alpha(j)$ is the largest rational below
$\alpha$ with denominator at most $j$.

\begin{lemma}
\label[lemma]{lem:computable-parameter}
For every computable $\alpha\in\mathcal A$,
$\CSP(\struct{B}_\alpha)$ is decidable.
\end{lemma}

\begin{proof}
Since $\alpha$ is computable and irrational, comparison with any rational is decidable by refining an approximation to $\alpha$ until its error interval excludes that rational. Choose an integer $c>\alpha$. For each $j$, a finite search over fractions $a/b$ with $0\le a\le cj$ and $1\le b\le j$ therefore computes $r_\alpha(j)$ exactly, and hence $S_\alpha(n)$ is an exactly computable rational. A finite graph $\struct{G}$ belongs to $\mathcal C_\alpha$ precisely when $e(\struct{H})\le S_\alpha(|\struct{H}|)$ for every nonempty induced $\struct{H}\subseteq\struct{G}$. These finitely many rational tests decide age membership. Enumerating the consistent quotients in \Cref{lem:quotient-completion} then decides $\CSP(\struct{B}_\alpha)$.
\end{proof}

\begin{theorem}[A polynomial-hierarchy upper bound]
\label[theorem]{thm:sigma-two-upper-bound}
Let $\alpha\in\mathcal A$. Suppose that there are an integer $c>\alpha$, a polynomial $p$, and a deterministic algorithm which, given $N\in\NN$ in unary and integers $a,b$ such that
\[
  0\leq a\leq cN
  \qquad\text{and}\qquad
  1\leq b\leq N,
\]
decides whether $a/b<\alpha$ in time at most $p(N)$. Then the restriction of $\CSP(\struct{B}_\alpha)$ to complete instances belongs to $\mathrm{coNP}$, and $\CSP(\struct{B}_\alpha)\in\Sigma_2^{\mathrm P}$.
\end{theorem}

\begin{proof}
Let $N$ be the number of vertices in the input instance. For each integer $j$ with $1\le j\le N$, enumerating the fractions $a/b$ with $0\le a\le cj$ and $1\le b\le j$, and using the assumed comparison algorithm, one can compute $r_\alpha(j)$ in time polynomial in the numerical value of $j$. Since $j\le N$, this bound is polynomial in the input size. Thus, all values $S_\alpha(k)$ for $k\le N$ can be computed exactly in polynomial time in $N$: their denominators divide $L_N\coloneqq\operatorname{lcm}(1,\ldots,N)\le N!$, which has $O(N\log N)$ bits, and their numerators also have polynomial bit length since $S_\alpha(k)<cN$.

Non-membership of a finite graph $\struct{G}$ in $\mathcal C_\alpha$ is certified by a nonempty subgraph $\struct{H}$ with $e(\struct{H})>S_\alpha(|\struct{H}|)$. This certificate is verifiable in polynomial time. By \Cref{lem:complete-age}, the complete-instance restriction therefore belongs to $\mathrm{coNP}$.

Now let $\struct{X}$ be an arbitrary instance of $\CSP(\struct{B}_\alpha)$. By \Cref{lem:quotient-completion}, we have
$\struct{X}\to\struct{B}_\alpha$ if and only if there is a consistent equivalence relation $\sim$ such that $\struct{G}_\sim\in\mathcal C_\alpha$, where $\struct{G}_\sim$ has the quotient classes as vertices and exactly the edges required by $E$. The relation $\sim$ has a polynomial-size encoding, and both consistency and the construction of $\struct{G}_\sim$ can be checked in polynomial time. The remaining condition is
\[
  e(\struct{S})\le S_\alpha(|\struct{S}|)
  \qquad\text{for every nonempty induced }
  \struct{S}\subseteq\struct{G}_\sim.
\]
Hence membership of $\struct{X}$ in $\CSP(\struct{B}_\alpha)$ has the form
\[
  \exists \sim \; \forall\,\struct{S}\subseteq\struct{G}_\sim\quad
  R(\struct{X},\sim,\struct{S}),
\]
where $R$ checks consistency and the inequality $e(\struct{S})\le S_\alpha(|\struct{S}|)$. These checks take polynomial time, so $\CSP(\struct{B}_\alpha)\in\Sigma_2^{\mathrm P}$.
\end{proof}

\begin{corollary}[An explicit upper-bound parameter]
\label[corollary]{cor:explicit-upper-bound}
There is an explicit irrational $\alpha\in\mathcal A$, with $9<\alpha<10$, for which the complete-instance restriction of $\CSP(\struct{B}_\alpha)$ belongs to $\mathrm{coNP}$ and $\CSP(\struct{B}_\alpha)\in\Sigma_2^{\mathrm P}$.
\end{corollary}

\begin{proof}
The parameter constructed in \Cref{prop:pantano-explicit-parameter} satisfies the comparison hypothesis of \Cref{thm:sigma-two-upper-bound} with $c=10$.
\end{proof}

Mere computability of $\alpha$ does not imply the polynomial-time comparison hypothesis in \Cref{thm:sigma-two-upper-bound}; in particular, we do not claim these upper bounds for the diagonal parameters constructed below.

\subsection{Effective parameters and a time hierarchy}
To prove \Cref{thm:hierarchy-intro} 
we select a computable parameter $\alpha$ through nested rational intervals.
We need to make the dense interval refinements effective;
decidability of $\CSP(\struct{B}_{\alpha})$ will then follow from
\Cref{lem:computable-parameter}.

\begin{lemma}[Finite parameter stratification]
\label[lemma]{lem:finite-stratification}
Let $r\in\NN$ and let $[a,b]\subseteq(1,\infty)$ have rational endpoints $a<b$. One can effectively compute a finite set $Q\subseteq[a,b]\cap\QQ$ such that, on every component of $[a,b]\setminus Q$, all functions
$\alpha\longmapsto F_\alpha(m)$ for $1\leq m\leq r$ 
are \emph{rational-affine}, i.e., affine with rational coefficients. Their formulae are computable from the component.
\end{lemma}

The proof is given in \Cref{app:effective-details}.

\begin{lemma}[Effective dense refinement]\label[lemma]{lem:effective-dense}
Given $m\in\NN$ and a nonempty open rational interval $I\subseteq(1,\infty)$, one can effectively find $K$ and a nonempty open rational interval $J$ with $\overline J\subseteq I$ such that
\[
  \sum_{N=1}^K\eta(\alpha,N)>m
  \qquad\text{for all }\alpha\in J.
\]
\end{lemma}

\begin{proof}
Enumerate all tuples $(K,c,d,c',d')$ with $K\geq1$ and rational numbers $c<c'<d'<d$ such that $[c,d]\subseteq I$. For each tuple, apply \Cref{lem:finite-stratification} with $r=K$ to $[c,d]$, obtaining a finite set $Q$. Check whether $[c',d']$ lies in a single component of $[c,d]\setminus Q$. If it does, then $F_\alpha(K)$ has a computable rational-affine formula on that component, hence, the minimum on $[c',d']$ occurs at an endpoint. Check whether its values at both $c'$ and $d'$ exceed $m$. If so, return $J\coloneqq(c',d')$ and $K$.

This search terminates: By \Cref{lem:evans-uniform-density}, there are $K$ and a nonempty open interval $J_0$ with $\overline{J_0}\subseteq I$ such that $F_\alpha(K)>m$ for every $\alpha\in J_0$. Choose rationals
$c<d$ with $[c,d]\subseteq I$ such that $(c,d)\cap J_0\neq\emptyset$. The finite set $Q$ leaves a component that intersects with $J_0$ in a nonempty open interval. We can choose rationals $c'<d'$ with $[c',d']$ inside that intersection. The tuple $(K,c,d,c',d')$ therefore passes both checks: $[c',d']$ lies in a single component of $[c,d]\setminus Q$, and the endpoint values $F_{c'}(K)$ and $F_{d'}(K)$ both exceed $m$. Since this tuple is eventually enumerated, the search terminates.
\end{proof}

\begin{proof}[Proof of \Cref{thm:hierarchy-intro}]
Fix an effective enumeration $(Q_s,c_s)_{s\in\NN}$ of all pairs consisting of a deterministic machine and a positive integer. Let $P_s$ simulate $Q_s$ for at most $c_s(t(n)+1)$ steps on an input with $n$ variables, rejecting if the clock expires. Since $t$ is time-constructible, these machines are effective and always halt. Also enumerate all rationals as $(r_s)_{s\in\NN}$.
For $s \in {\mathbb N}$, we construct nonempty rational open intervals $I_s$ and finite graphs $\struct{G}_s$ satisfying:
\begin{enumerate}[(i)]
\item $|I_s|\le2^{-s}$, and $\overline{I_s}\subseteq I_{s-1}$ for $s>0$;
\item\label{item:avoidrationals}
$\overline{I_s}\subseteq U_s$ (where $U_s$ is the open dense set defined in Section~\ref{sect:admiss}) and it does not contain $r_0,\ldots,r_{s-1}$;
\item\label{item:disagree_with_P_s}
for every $j<s$ and $\beta\in I_s\cap\mathcal A$,
\[
  \struct{G}_j\in\mathcal C_\beta
  \quad\Longleftrightarrow\quad
  P_j\text{ rejects }\struct{X}_{\struct{G}_j}.
\]
\end{enumerate}

Choose $I_0$ with $\overline{I_0}\subseteq(1,\infty)$ and $|I_0|\le1$. Since $F_\beta(1)>0$ for $\beta>1$, its closure lies in $U_0$; the avoidance and disagreement requirements in (ii) and (iii) are initially vacuous.

Given $I_s$, apply \Cref{lem:effective-dense} with threshold $m :=s+1$ to obtain an interval $I'_s$ with $\overline{I'_s}\subseteq I_s$ and $I'_s\subseteq U_{s+1}$. Shrink it so that $r_s\notin\overline{I'_s}$.
Choose nonempty rational open intervals $J_s^0,J_s^1$ whose closures lie in $I'_s$ and satisfy $\sup J_s^0<\inf J_s^1$. The effective construction following \Cref{lem:switch-main} gives a finite graph $\struct{G}_s$ such that $\struct{G}_s\notin\mathcal C_\beta$ for $\beta\in J_s^0\cap\mathcal A$, whereas $\struct{G}_s\in\mathcal C_\beta$ for $\beta\in J_s^1\cap\mathcal A$.

Run $P_s$ on $\struct{X}_{\struct{G}_s}$ and set
\[
  K_s:=
  \begin{cases}
    J_s^0,&\text{if }P_s\text{ accepts }
              \struct{X}_{\struct{G}_s},\\
    J_s^1,&\text{if }P_s\text{ rejects }
              \struct{X}_{\struct{G}_s}.
  \end{cases}
\]
Choose a nonempty rational open interval $I_{s+1}$ with $\overline{I_{s+1}}\subseteq K_s$ and $|I_{s+1}|\le2^{-(s+1)}$. Its closure avoids $r_0,\ldots,r_s$ and lies in $U_{s+1}$. Since $I_{s+1}\subseteq I_s$, for $\beta \in I_{s+1}\cap \mathcal{A}$ and $j<s$ it still holds that $\struct{G}_j\in\mathcal C_\beta$ if and only if $P_j$ rejects $\struct{X}_{\struct{G}_j}$. The choice of $K_s$ ensures that the same equivalence holds for $j=s$.

The nested compact intervals $\overline{I_s}$ have diameters tending to zero. Define their unique common point by
\[
  \{\alpha\}:=\bigcap_s\overline{I_s}.
\]
Their effectively computed endpoints approximate $\alpha$ with error at most $2^{-s}$, so $\alpha$ is computable. Requirement~\eqref{item:avoidrationals} ensures that $\alpha$ is irrational and belongs to every $U_s$, hence $\alpha\in\mathcal A$. By \Cref{prop:tameness} and \Cref{rem:definable-expansions}, $\struct{M}_\alpha$ and its definable expansion $\struct{B}_\alpha$ are stable and $\omega$-categorical. By \Cref{lem:computable-parameter}, $\CSP(\struct{B}_\alpha)$ is decidable.

Since $\overline{I_{s+1}}\subseteq I_s$, the parameter $\alpha$ belongs to every $I_s$. Requirement~\eqref{item:disagree_with_P_s} and \Cref{lem:complete-age} therefore show that $P_s$ gives the wrong answer for $\CSP(\struct{B}_\alpha)$ on $\struct{X}_{\struct{G}_s}$, for every $s$.

Suppose that a deterministic machine $Q$ correctly decides all complete instances in time $O(t(n))$. Choose a positive integer $c$ such that $Q$ halts within $c(t(n)+1)$ steps on every complete instance. The pair $(Q,c)$ occurs as $(Q_s,c_s)$ for some $s$. On $\struct{X}_{\struct{G}_s}$, the clocked machine $P_s$ then agrees with $Q$, a contradiction. Thus the lower bound holds already on complete instances, and $\CSP(\struct{B}_\alpha)\notin\DTIME(t)$.
\end{proof}

\section{CSP complexity of the graph reducts}
\label{sec:colouring-complexity}
The age separation above does not separate the CSPs of the pure graph reducts: a graph homomorphism need not preserve non-edges. We now analyse the CSPs for the pure graphs $\struct{M}_\alpha$. Write $\struct{K}_s$ for the complete graph on $s$ vertices. We call $\struct{G}$ and $\struct{H}$ homomorphically equivalent, written $\struct{G}\longleftrightarrow\struct{H}$,
if $\struct{G}\to\struct{H}$ and $\struct{H}\to\struct{G}$. Homomorphically equivalent graphs have the same CSP, even when their ages differ.

We prove \NP-hardness of $\CSP(\struct{M}_\alpha)$ (and hence, of $\CSP(\struct{B}_\alpha)$) for every admissible irrational $\alpha>1$ using hardness results for promise graph colouring. For $\alpha\in(1,5/3)\cup(2,\infty)$, we show that $\CSP(\struct{M}_\alpha)$ is equivalent to $k$-colouring for some $k\geq 3$ and therefore \NP-complete. For parameters in $\mathcal A\cap\bigl((1,5/3)\cup(2,\infty)\bigr)$, continuum many templates therefore yield exactly countably many distinct CSP languages, all \NP-complete.

Throughout this subsection, $\alpha>1$ is assumed to be
admissible and irrational. By \eqref{eq:lower-approximations},
inequality~\eqref{eq:control} for a finite graph $\struct{H}$
is equivalent to $e(\struct{H})\le S_\alpha(|\struct{H}|)$.
Every finite subgraph of $\struct{M}_\alpha$ satisfies
this inequality.

\begin{theorem}
\label[theorem]{thm:graph-hardness}
For every admissible irrational $\alpha>1$, with $k:=\lfloor\alpha\rfloor$,
\[
  \struct{K}_{2k+1}\longrightarrow\struct{M}_\alpha
  \longrightarrow\struct{K}_{2k+2}.
\]
Hence, both $\CSP(\struct{M}_\alpha)$ and
$\CSP(\struct{B}_\alpha)$ are \NP-hard.
\end{theorem}

\begin{proof}
Since $r_\alpha(j)\ge k$ for all $j$, we have $S_\alpha(s)\ge ks\ge\binom{s}{2}$ for
$1\le s\le2k+1$. Thus, $\struct{K}_{2k+1}\in\mathcal C_\alpha$, giving the first homomorphism.

Every nonempty finite subgraph $\struct{H}$ of $\struct{M}_\alpha$ satisfies $\alpha|\struct{H}|-e(\struct{H})>0$, hence, has average degree $2e(\struct{H})/|\struct{H}|<2\alpha<2k+2$. Repeatedly deleting a vertex of degree at most $2k+1$ and then colouring in reverse order (the previously coloured neighbours forbid at most $2k+1$ colours) gives a $(2k+2)$-colouring. Compactness therefore yields $\struct{M}_\alpha\to\struct{K}_{2k+2}$.

Put $\ell:=2k+1\ge3$. By~\cite[Theorem~6.5]{BBKO}, distinguishing $\ell$-colourable graphs from graphs that are not $(2\ell-1)$-colourable is \NP-hard. The chain
\[
  \struct{K}_\ell\longrightarrow\struct{M}_\alpha
  \longrightarrow\struct{K}_{2k+2}
  \longrightarrow\struct{K}_{2\ell-1}
\]
shows that an algorithm for $\CSP(\struct{M}_\alpha)$ distinguishes these cases. Adding an empty $N$-relation to an input reduces $\CSP(\struct{M}_\alpha)$ to $\CSP(\struct{B}_\alpha)$, hence, $\CSP(\struct{B}_\alpha)$ is also NP-hard. 
\end{proof}

\begin{proposition}
\label[proposition]{prop:graph-intervals}
For every admissible irrational $\alpha>1$, put $k:=\lfloor\alpha\rfloor$ and $\beta_k:=k+\frac12+\frac{1}{4\lfloor k/2\rfloor+2}$. The following homomorphic equivalences hold:
\[
  \struct{M}_\alpha\longleftrightarrow
  \begin{cases}
    \struct{K}_3,
      &1<\alpha<5/3,\\
    \struct{K}_{2k+1},
      &k\ge2,\ \alpha<\beta_k,\\
    \struct{K}_{2k+2},
      &k\ge2,\ \alpha>\beta_k.
  \end{cases}
\]
Hence, $\CSP(\struct{M}_\alpha)$ is \NP-complete for $\alpha\in(1,5/3)\cup(2,\infty)$.
\end{proposition}

\begin{proof}[Proof sketch; full proof in \Cref{app:colouring-details}]
For $1<\alpha<5/3$, the estimates
$r_\alpha(1)=1$, $r_\alpha(2)\le3/2$, and
$r_\alpha(j)<5/3$ give
$S_\alpha(n)<(5n-2)/3$ for $n\ge4$.
The Kostochka--Yancey bound
\cite[Theorem~4]{KYdensity} and compactness yield
$\struct{M}_\alpha\to\struct{K}_3$. Together with \Cref{thm:graph-hardness}, this gives $\struct{M}_\alpha\longleftrightarrow\struct{K}_3$.

For $\alpha>2$, put $\ell=2k+2$.
If $S_\alpha(\ell)\ge\binom{\ell}{2}$, then
$\struct{K}_\ell$ embeds into $\struct{M}_\alpha$.
Otherwise, \Cref{lem:colouring-estimates} and \cite[Theorem~4]{KYdensity} give $\struct{M}_\alpha\to\struct{K}_{\ell-1}$.
\Cref{thm:graph-hardness} yields $\struct{K}_{\ell-1}\to\struct{M}_\alpha
  \to\struct{K}_{\ell}$, hence, $\struct{M}_\alpha\longleftrightarrow\struct{K}_{\ell}$ in the first case and $\struct{M}_\alpha\longleftrightarrow\struct{K}_{\ell-1}$ in the second. Finally, \Cref{lem:clique-cutoffs} gives
the thresholds $\beta_k$.
\end{proof}

Put $\mathcal A_0:=\mathcal A\cap\bigl((1,5/3)\cup(2,\infty)\bigr)$.
By \Cref{prop:graph-intervals},
\[
  \bigl\{\CSP(\struct{M}_\alpha):\alpha\in\mathcal A_0\bigr\}
  =
  \bigl\{\CSP(\struct{K}_s):s=3\text{ or }s\ge5\bigr\},
\]
and by \Cref{lem:comeager-continuum} each language on the right is realised by continuum many parameters. Thus, continuum many graph templates with pairwise distinct ages by \Cref{lem:switch-main} yield exactly countably many distinct CSPs, all \NP-complete. For $\alpha\in\mathcal A\cap(5/3,2)$, \Cref{thm:graph-hardness} gives $\struct{K}_3\to\struct{M}_\alpha\to\struct{K}_4$. Thus, $\CSP(\struct{M}_\alpha)$ is \NP-hard, while membership in NP remains open.

The expansions $\struct{B}_\alpha$, for $\alpha\in\mathcal A$, have pairwise distinct CSPs by \Cref{lem:complete-age}, and all but countably many are undecidable. Consequently, for all but countably many
$\alpha\in\mathcal A_0$, adding the non-edge relation changes an \NP-complete CSP into an undecidable one, while preserving stability and $\omega$-categoricity.

\begin{proof}[Proof of \Cref{cor:three-regimes}]
Apply the construction in the proof of \Cref{thm:hierarchy-intro} with $t(n)\coloneqq2^{2^n}$ and an initial rational interval whose closure lies in $(1,5/3)\cup(2,\infty)$. This yields a computable irrational $\alpha\in\mathcal A\cap\bigl((1,5/3)\cup(2,\infty)\bigr)$ such that $\CSP(\struct{B}_\alpha)$ is decidable but does not belong to $\DTIME(t)$, with time measured by the number of variables. In a fixed finite binary signature, an instance on $n$ variables admits an encoding of size $O(n^2)$. Membership in NP would therefore give a deterministic $2^{n^{O(1)}}$-time algorithm, and hence membership in $\DTIME(t)$. Thus $\CSP(\struct{B}_\alpha)$ is decidable but not in NP.

For the same $\alpha$, \Cref{prop:graph-intervals} gives that $\CSP(\struct{M}_\alpha)$ is NP-complete. Both templates are countably infinite, stable and $\omega$-categorical, with one and two binary relations, respectively. Finally, \Cref{thm:undecidable-intro} supplies the undecidable case.
\end{proof}

\section{Conclusion}
We construct continuum many countable stable $\omega$-categorical graphs with pairwise distinct ages. Their expansions by the non-edge relation have pairwise distinct CSPs and all but countably many of them are undecidable. We also obtain computable parameters such that the corresponding CSPs are decidable but 
the running time of any deterministic algorithm that solves such a CSP 
exceeds any prescribed time-constructible time bound. Thus, stability and $\omega$-categoricity alone impose no uniform bound on CSP complexity. 
For the graph reducts, we prove NP-hardness for every admissible irrational $\alpha>1$ and NP-completeness for $\alpha\in(1,5/3)\cup(2,\infty)$. In particular, for all but countably many $\alpha$ in this range, adding the non-edge relation changes an NP-complete CSP into an undecidable one. For suitable computable $\alpha$, the expansion instead has a decidable CSP outside NP. The interval $(5/3,2)$ remains to be understood. There, our results establish NP-hardness of $\CSP(\struct{M}_\alpha)$, but do not establish membership in NP.

Our cardinality argument does not identify an explicit parameter $\alpha$ for which $\CSP(\struct{B}_\alpha)$ is undecidable, and the computable parameters with time lower bounds are obtained by diagonalisation. For Pantano's explicit parameter, the restriction to complete instances belongs to coNP and the full CSP belongs to $\Sigma_2^{\mathrm P}$. It remains to determine the exact complexity in both cases and whether the passage from complete to arbitrary instances increases the complexity.

Our non-dichotomy relies on CSPs outside NP. It remains open whether stable $\omega$-categorical templates can have NP-intermediate CSPs. We ask whether such an example can be found among the first-order reducts of the graphs $\struct{M}_\alpha$. A positive answer would also answer Question~\ref{quest:NP-intermediate} positively.

\begin{question}
Is there an admissible irrational $\alpha>1$ and a first-order reduct $\struct{C}$ of $\struct{M}_\alpha$ such that $\CSP(\struct{C})$ is NP-intermediate?
\end{question}

\bibliography{hrushovski}

\newpage
\appendix

\section{Construction of the amalgamation limit}\label{app:generic-construction}

\begin{proof}[Proof of \Cref{thm:am_lim}]
Enumerate representatives of the isomorphism types in $\mathcal K$ as $\struct{A}_0,\struct{A}_1,\ldots$. We construct a chain
\[
  \struct{M}_0\leq\struct{M}_1\leq\struct{M}_2\leq\cdots
\]
of members of $\mathcal K$. Start with any $\struct{M}_0\in\mathcal{K}$.

Suppose that $\struct{M}_n$ has been constructed. By the joint embedding property, choose a strong extension $\struct{C}$ of $\struct{M}_n$ into which $\struct{A}_n$ strongly embeds.

For every $j\leq n$, every $\struct{D}\leq\struct{A}_j$, and every strong embedding $f\colon\struct{D}\to\struct{M}_n$, successively amalgamate the current $\struct{C}$ and $\struct{A}_j$ over $\struct{D}$, using $f$ and the inclusion into $\struct{A}_j$.
Transitivity ensures that $f$ is strong as a map into the current $\struct{C}$.
Replacing the amalgam by an isomorphic copy, we retain $\struct{C}$ as a strong substructure and obtain a strong embedding of $\struct{A}_j$ extending $f$.
There are only finitely many such embeddings to process, since the structures involved are finite.
After these amalgamations, let $\struct{M}_{n+1}$ be the resulting structure.
Transitivity ensures that all embeddings obtained during the stage remain strong in $\struct{M}_{n+1}$.

Let $\struct{M}\coloneqq\bigcup_n\struct{M}_n$. This structure is countable, and heredity gives
$\Age(\struct{M})\subseteq\mathcal K$.
Moreover, $\struct{M}_i\leq\struct{M}$ for every $i$: given a finite substructure $\struct{D}\subseteq\struct{M}$ containing $\struct{M}_i$, choose $k\geq i$ with $\struct{D}\subseteq\struct{M}_k$.
Transitivity gives $\struct{M}_i\leq\struct{M}_k$, and convexity gives $\struct{M}_i\leq\struct{D}$.
Since every finite subset of $M$ lies in some $M_i$, this proves~\eqref{finitesubsetcontainedinstrongsubstr}.

At stage $j$, $\struct{A}_j$ strongly embeds into $\struct{M}_{j+1}$, and hence into $\struct{M}$.
Thus, every member of $\mathcal K$ strongly embeds into $\struct{M}$, proving~\eqref{ageamalgamationlimit}.
Since $\mathcal K$ contains structures of arbitrarily large finite size, $\struct{M}$ is countably infinite.

To prove~\eqref{extensionamalgamationlimit}, let $\struct{A}\leq\struct{B}\in\mathcal K$ and let $f\colon\struct{A}\to\struct{M}$ be strong.
Up to isomorphism, we may assume that $\struct{B}=\struct{A}_j$ for some $j$.
Choose $n\geq j$ with $f(A)\subseteq M_n$.
Since $f(\struct{A})$ is strong in $\struct{M}$, it is strong in $\struct{M}_n$.
The construction at stage $n$ therefore supplies a strong embedding of $\struct{B}$ into $\struct{M}_{n+1}$ extending $f$.
Its image is strong in $\struct{M}$, as required.
If $\struct{M}$ and $\struct{M}'$ both have these properties, the isomorphisms between finite strong substructures form a back-and-forth system. Alternating extensions along enumerations of $M$ and $M'$ gives an isomorphism. Starting with an isomorphism between finite strong substructures of one copy gives homogeneity restricted to strong substructures.
\end{proof}

\begin{proof}[Proof of \Cref{lem:bounded-strong-substructures}]
Replacing $b(n)$ by $\max_{i\le n}b(i)$, we may assume that $b$ is non-decreasing. Fix $n$, and let $\bar a$ be an $n$-tuple from the amalgamation limit $\struct{M}$.
Write $A$ for the set of entries of $\bar a$. By \Cref{thm:am_lim}(ii), there is a finite $\struct{B}\leq\struct{M}$ containing $A$. The hypothesis applied inside $\struct{B}\in\mathcal K$ gives a strong $\struct{C}\leq\struct{B}$ containing $A$ with $|\struct{C}|\leq b(n)$. The observation preceding \Cref{thm:am_lim} gives $\struct{C}\leq\struct{M}$. Thus, every $n$-tuple is contained in a finite strong substructure of size at most $b(n)$. There are only finitely many isomorphism types of pairs $(\struct{C},\bar a)$ with $|\struct{C}|\leq b(n)$ and $\bar a\in C^n$. Homogeneity with respect to strong substructures shows that if two tuples lie in isomorphic such pairs, they lie in the same orbit. Hence, $\Aut(\struct{M})$ has finitely many orbits on $M^n$ for every $n$, and Ryll-Nardzewski--Engeler--Svenonius gives $\omega$-categoricity.
\end{proof}

\section{Controlled predimension calculations}
\label{app:controlled-details}

We collect the standard finite calculations used in
\Cref{lem:approximation-estimates,lem:strongness-calculus,lem:amalgamation,lem:closure-predimension}.
They are included to make explicit the specialisation of the controlled
predimension construction to the present graph parameters.

\begin{proof}[Proof of \Cref{lem:approximation-estimates}]
The minimum defining $\eta(\alpha,N)$ is taken over a finite nonempty set and
is positive because every candidate is required to be strictly below
$\alpha$. Increasing $N$ enlarges the set of rational approximations, proving
monotonicity. Taking $a\coloneqq \lceil N\alpha\rceil-1$ gives
\[
  0<\alpha-\frac{a}{N}\leq\frac1N,
\]
and therefore $\eta(\alpha,N)\leq1/N$. The slope of $F_\alpha$ on
$[N,N+1]$ is $\eta(\alpha,N+1)$. These slopes are positive and
non-increasing, which proves strict increase and concavity. Summing the
upper bounds gives $F_\alpha(N)\leq \sum_{k=1}^{N}\frac1k$, and $(\sum_{k=1}^N\frac1k)/N\to0$.
\end{proof}

\begin{proof}[Proof of \Cref{lem:strongness-calculus}]
Reflexivity is immediate. If $\struct{A}\leq_\alpha\struct{B}\leq_\alpha\struct{C}$ and
$\struct{A}\subsetneq\struct{X}\subseteq\struct{C}$, then either $\struct{X}\subseteq\struct{B}$, in which case the
claim follows from $\struct{A}\leq_\alpha\struct{B}$, or $\struct{B}\subsetneq\struct{B}\cup\struct{X}\subseteq\struct{C}$.
In the latter case, $\struct{B}\leq_\alpha\struct{C}$ and \eqref{eq:submod} give
\[
  \delta_\alpha(\struct{X})-\delta_\alpha(\struct{B}\cap\struct{X})
  \geq
  \delta_\alpha(\struct{B}\cup\struct{X})-\delta_\alpha(\struct{B})>0.
\]
Since $\struct{A}\subseteq\struct{B}\cap\struct{X}\subseteq\struct{B}$, either $\struct{B}\cap\struct{X}=\struct{A}$ or
$\delta_\alpha(\struct{B}\cap\struct{X})>\delta_\alpha(\struct{A})$. Thus
\[
  \delta_\alpha(\struct{X})>\delta_\alpha(\struct{B}\cap\struct{X})\geq\delta_\alpha(\struct{A}),
\]
proving transitivity.

For the intersection assertion, let
$\struct{A}\cap\struct{X}\subsetneq\struct{Y}\subseteq\struct{X}$. Then $\struct{A}\subsetneq\struct{A}\cup\struct{Y}\subseteq\struct{B}$,
and \eqref{eq:submod} together with $\struct{A}\leq_\alpha\struct{B}$ gives
\[
  \delta_\alpha(\struct{Y})-\delta_\alpha(\struct{A}\cap\struct{X})
  \geq \delta_\alpha(\struct{A}\cup\struct{Y})-\delta_\alpha(\struct{A})>0.
\]
Hence $\struct{A}\cap\struct{X}\leq_\alpha\struct{X}$.
\end{proof}

\begin{proof}[Proof of \Cref{lem:amalgamation}]
The class is hereditary by definition, it is closed under isomorphisms, and has
only countably many isomorphism types. The empty graph belongs to
$\mathcal C_\alpha$, since the defining condition quantifies only over
nonempty induced subgraphs. Moreover, $\varnothing\leq_\alpha\struct{B}$ for every
$\struct{B}\in\mathcal C_\alpha$, since every
nonempty induced $\struct{B}'\subseteq\struct{B}$ satisfies
$\delta_\alpha(\struct{B}')\geq F_\alpha(|\struct{B}'|)>0$. Thus, the joint embedding property
follows by amalgamating over the empty graph once amalgamation is established.

Let $\struct{A}\leq_\alpha\struct{B}_1,\struct{B}_2$ and let $\struct{D}$ be their free amalgam over $\struct{A}$.
We first prove inequality~\eqref{eq:control} for an arbitrary nonempty induced
$\struct{X}\subseteq\struct{D}$. Put
\[
  \struct{X}_i\coloneqq\struct{X}\cap\struct{B}_i \text{ for }i\in\{1,2\},
  \qquad
  \struct{X}_0\coloneqq\struct{X}\cap\struct{A}.
\]
Then $\struct{X}$ is the free amalgam of $\struct{X}_1$ and $\struct{X}_2$ over $\struct{X}_0$, and
\Cref{lem:strongness-calculus} gives $\struct{X}_0\leq_\alpha\struct{X}_i$. If one of the
$\struct{X}_i$ equals $\struct{X}_0$, then $\struct{X}$ is isomorphic to the other $\struct{X}_i$ and the claim
follows because $\struct{B}_i\in\mathcal C_\alpha$. Otherwise, possibly after
interchanging $\struct{X}_1$ and $\struct{X}_2$, suppose
\[
  \frac{\delta_\alpha(\struct{X}_2)-\delta_\alpha(\struct{X}_0)}{|\struct{X}_2|-|\struct{X}_0|}
  \geq
  \frac{\delta_\alpha(\struct{X}_1)-\delta_\alpha(\struct{X}_0)}{|\struct{X}_1|-|\struct{X}_0|}.
\]
The last quotient is positive because
$\struct{X}_0\subsetneq\struct{X}_1$ and $\struct{X}_0\leq_\alpha\struct{X}_1$. Write it as
\[
  \alpha-q,
  \qquad
  q\coloneqq\frac{e(\struct{X}_1)-e(\struct{X}_0)}{|\struct{X}_1|-|\struct{X}_0|}.
\]
The denominator of $q$ is at most $|\struct{X}_1|$, and $q<\alpha$. Therefore
\[
  \alpha-q\geq\eta(\alpha,|\struct{X}_1|).
\]
Set
\[
  x\coloneqq|\struct{X}_1|,
  \qquad
  y\coloneqq|\struct{X}_2|-|\struct{X}_0|.
\]
Thus, $x,y\geq1$ and $x+y=|\struct{X}|$. The assumed ordering yields
\begin{equation}\label{eq:amalgam-increment}
  \delta_\alpha(\struct{X}_2)-\delta_\alpha(\struct{X}_0)
  \geq y\eta(\alpha,x).
\end{equation}
Moreover, $\struct{X}_1\in\mathcal C_\alpha$, so
\begin{equation}\label{eq:amalgam-base}
  \delta_\alpha(\struct{X}_1)\geq F_\alpha(x).
\end{equation}
Concavity of the control function gives
\begin{equation}\label{eq:amalgam-concavity}
  F_\alpha(x+y)
  \leq F_\alpha(x)+y\eta(\alpha,x+1)
  \leq F_\alpha(x)+y\eta(\alpha,x).
\end{equation}
Since $\struct{X}$ is the free amalgam of $\struct{X}_1$ and $\struct{X}_2$ over $\struct{X}_0$, these bounds
combine to give
\begin{align*}
  \delta_\alpha(\struct{X})
  &=\delta_\alpha(\struct{X}_1)+\delta_\alpha(\struct{X}_2)-\delta_\alpha(\struct{X}_0)\\
  &\geq \delta_\alpha(\struct{X}_1)+y\eta(\alpha,x)\\
  &\geq F_\alpha(x)+y\eta(\alpha,x)\\
  &\geq F_\alpha(x+y)
  =F_\alpha(|\struct{X}|).
\end{align*}
Thus, $\struct{D}\in\mathcal C_\alpha$.

Finally, if $\struct{B}_1\subsetneq\struct{Y}\subseteq\struct{D}$, then
$\struct{A}\subsetneq\struct{Y}\cap\struct{B}_2\subseteq\struct{B}_2$, and freeness gives
\[
  \delta_\alpha(\struct{Y})-\delta_\alpha(\struct{B}_1)
  =\delta_\alpha(\struct{Y}\cap\struct{B}_2)-\delta_\alpha(\struct{A})>0.
\]
Hence $\struct{B}_1\leq_\alpha\struct{D}$, and symmetrically $\struct{B}_2\leq_\alpha\struct{D}$. Thus, the
free amalgam is an amalgam by strong embeddings.
\end{proof}

\begin{proof}[Proof of \Cref{lem:closure-predimension}]
Starting with $\struct{A}_0\coloneqq\struct{A}$, if $\struct{A}_i$ is not strong in $\struct{B}$, choose
$\struct{A}_i\subsetneq\struct{A}_{i+1}\subseteq\struct{B}$ with
$\delta_\alpha(\struct{A}_{i+1})\leq\delta_\alpha(\struct{A}_i)$. Finiteness of $\struct{B}$ makes
this process terminate at a strong $\struct{C}$ containing $\struct{A}$, with
$\delta_\alpha(\struct{C})\leq\delta_\alpha(\struct{A})$. The smallest strong closure
$\cl_{\struct{B}}(\struct{A})$ is contained in $\struct{C}$. If this containment is proper, strongness
of $\cl_{\struct{B}}(\struct{A})$ gives
$\delta_\alpha(\cl_{\struct{B}}(\struct{A}))<\delta_\alpha(\struct{C})$; otherwise they are equal.
\end{proof}

\section{Stability of the controlled amalgamation limit}\label{app:stability}
The stability argument given below is an instance of Wagner's general dimension-theoretic analysis of Hrushovski generics
\cite[Section~3 and Example~5.3]{Wagner94}, while a proof that is specialised for limits of classes of graphs controlled by the same function is written by Herwig \cite{Herwig95} (see the first paragraph of Section 6 and the proof of stability in Section 4 culminating in Theorem 14). It also closely parallels Pantano's proof for her controlled stable constructions in a countably infinite language \cite[Section~4.3, pp.~51--61]{Pantano1995}. We include the specialisation
needed here in order to verify explicitly the points required for the present
graph class and for arbitrary models of its complete theory. We prove stability by showing that types over a closed parameter set are determined by countable closed extensions of countable subsets of that set. We first establish the required closure and dimension properties and show that the extension property from \Cref{thm:am_lim}~(\ref{extensionamalgamationlimit}) holds in every model of $T_\alpha$. We then use dimension independence to obtain closed free amalgams, whose isomorphism types determine complete types. The resulting type count gives stability.

For an infinite cardinal $\lambda$, a complete theory $T$ is
\emph{$\lambda$-stable} if, for every parameter set $P$ of cardinality at
most $\lambda$ and every $n<\omega$, there are at most $\lambda$ complete
$n$-types over $P$. Equivalently, it suffices to count $1$-types. A theory
is stable in this type-counting sense if it is $\lambda$-stable for some
infinite $\lambda$. For complete first-order theories this is equivalent to
the definition of stability by the absence of the order property (the
definition recalled in \Cref{sec:preliminaries}), and due to Shelah
\cite[Theorem~II.2.13]{Shelah90}; also see for example
\cite[Section~5.2 and Exercise~8.2.7]{TZ12}. Shelah's foundational paper \cite{Shelah69} develops the
type-counting approach to stability and studies the stability
spectrum, that is, the infinite cardinals $\lambda$ for which
a theory is $\lambda$-stable. The proof below establishes $\lambda$-stability directly
for $\lambda=2^{\aleph_0}$. No novelty is claimed for the underlying
stability method.

\paragraph*{Closure and dimension.}
Fix an admissible irrational $\alpha>1$, i.e. $F_\alpha(N)\to\infty$, and put 
\[
  \partial(\struct{A})\coloneqq\alpha^{-1}\delta_\alpha(\struct{A})
  =|\struct{A}|-\alpha^{-1}e(\struct{A}),
  \qquad
  f(N)\coloneqq\alpha^{-1}F_\alpha(N).
\]
Thus, $\mathcal C_\alpha$ consists exactly of the finite graphs $\struct{A}$ such that
$\partial(\struct{B})\geq f(|\struct{B}|)$ for every nonempty induced $\struct{B}\subseteq\struct{A}$.
Positive rescaling does not change the strong-substructure relation.
For finite induced subgraphs $\struct{U},\struct{V}$ of a common graph, direct edge counting gives
\[
  e(\struct{U})+e(\struct{V})\leq e(\struct{U}\cup\struct{V})+e(\struct{U}\cap\struct{V}),
\]
and hence
\begin{equation}\label{eq:appendix-submodularity}
  \partial(\struct{U}\cup\struct{V})+\partial(\struct{U}\cap\struct{V})
  \leq\partial(\struct{U})+\partial(\struct{V}).
\end{equation}

Let $\struct{N}$ be a graph all of whose finite induced subgraphs belong to
$\mathcal C_\alpha$. For finite $\struct{X}\subseteq\struct{N}$, define
\[
  d_{\struct{N}}(\struct{X})\coloneqq
  \min\{\partial(\struct{Y}):\struct{X}\subseteq\struct{Y}\subseteq\struct{N}, \struct{Y}\text{ finite}\}.
\]
The minimum exists. Indeed, it is enough to consider $\struct{Y}$ with
$\partial(\struct{Y})\leq\partial(\struct{X})$, and then
$f(|\struct{Y}|)\leq\partial(\struct{X})$ bounds $|\struct{Y}|$ because $f(N)\to\infty$. Hence only
finitely many values of $\partial(\struct{Y})$ can occur. Submodularity shows that
the intersection of two minimisers is again a minimiser: if both have value
$r$, then their union and intersection have value at least $r$, while
submodularity forces both values to equal $r$. We write $\cl_{\struct{N}}(\struct{X})$ for the
intersection of all minimisers. It is finite and
\begin{equation}\label{eq:appendix-dimension-closure}
  d_{\struct{N}}(\struct{X})=\partial(\cl_{\struct{N}}(\struct{X})).
\end{equation}
Since $\alpha$ is irrational, a proper finite extension cannot have the same
$\partial$-value. Consequently a finite $\struct{X}$ is closed in $\struct{N}$ precisely when
$d_{\struct{N}}(\struct{X})=\partial(\struct{X})$, and then $\cl_{\struct{N}}(\struct{X})=\struct{X}$.

If $\struct{X}\subseteq\struct{Y}$ are finite, put
$\struct{C}\coloneqq\cl_{\struct{N}}(\struct{X})$ and $\struct{D}\coloneqq\cl_{\struct{N}}(\struct{Y})$. The sets $\struct{C}\cap\struct{D}$ and
$\struct{C}\cup\struct{D}$ contain $\struct{X}$ and $\struct{Y}$, respectively, so their $\partial$-values are at
least $d_{\struct{N}}(\struct{X})$ and $d_{\struct{N}}(\struct{Y})$. Applying
\eqref{eq:appendix-submodularity} to $\struct{C},\struct{D}$ shows that equality holds
throughout. Thus, $\struct{C}\cap\struct{D}$ is a minimiser for $\struct{X}$; by the definition of $\struct{C}$
as the intersection of all such minimisers,
$\struct{C}\subseteq\struct{C}\cap\struct{D}\subseteq\struct{D}$. Hence finite closure is monotone. Moreover,
$\cl_{\struct{N}}(\struct{X})$ is closed: every finite set containing it is also a candidate in the
minimum defining $d_{\struct{N}}(\struct{X})$ and therefore has $\partial$-value at least
$\partial(\cl_{\struct{N}}(\struct{X}))$.

For arbitrary $\struct{Z}\subseteq\struct{N}$, set
\[
  \cl_{\struct{N}}(\struct{Z})\coloneqq
  \bigcup\{\cl_{\struct{N}}(\struct{Z}_0):\struct{Z}_0\subseteq\struct{Z}\text{ finite}\}.
\]
We show that $\cl_{\struct{N}}$ is a finitary closure operator. By construction, $\cl_{\struct{N}}$ has finite character: the closure of a set is the union of the closures of its finite subsets. It is also extensive, i.e.  $\struct{Z}\subseteq\cl_{\struct{N}}(\struct{Z})$, and monotone. To show idempotence, i.e. $\cl_{\struct{N}}(\cl_{\struct{N}}(\struct{Z}))=\cl_{\struct{N}}(\struct{Z})$, note that for $\struct{W}\subseteq\cl_{\struct{N}}(\struct{Z})$
finite, we have $\struct{W}\subseteq\cl_{\struct{N}}(\struct{Z}_0)$ for some finite $\struct{Z}_0\subseteq\struct{Z}$. Finite monotonicity and idempotence give
$\cl_{\struct{N}}(\struct{W})\subseteq\cl_{\struct{N}}(\cl_{\struct{N}}(\struct{Z}_0))=\cl_{\struct{N}}(\struct{Z}_0)\subseteq\cl_{\struct{N}}(\struct{Z})$. It follows that $\cl_{\struct{N}}(\cl_{\struct{N}}(\struct{Z}))=\cl_{\struct{N}}(\struct{Z})$, so $\cl_{\struct{N}}(\struct{Z})$ is closed.

We establish four properties of dimension that will be used below.
Throughout, $\struct{X},\struct{Y},\struct{U},\struct{V}$ are finite subsets of $\struct{N}$.
\begin{enumerate}[(D1)]
\item \textbf{Non-negativity, monotonicity, and submodularity.}
We have $d_{\struct{N}}(\struct{X})\geq0$, and $\struct{X}\subseteq\struct{Y}$ implies
$d_{\struct{N}}(\struct{X})\leq d_{\struct{N}}(\struct{Y})$. Moreover,
\[
  d_{\struct{N}}(\struct{X}\cup\struct{Y})+d_{\struct{N}}(\struct{X}\cap\struct{Y})
  \leq d_{\struct{N}}(\struct{X})+d_{\struct{N}}(\struct{Y}).
\]
\end{enumerate}
\begin{proof}
Non-negativity follows because every finite graph considered in the
definition of $d_{\struct{N}}$ belongs to $\mathcal C_\alpha$. Monotonicity
follows because enlarging $\struct{X}$ restricts the finite supersets over
which the minimum is taken. For submodularity, apply
\eqref{eq:appendix-submodularity} to $\cl_{\struct{N}}(\struct{X})$ and $\cl_{\struct{N}}(\struct{Y})$.
Their union and intersection are candidates in the definitions of
$d_{\struct{N}}(\struct{X}\cup\struct{Y})$ and $d_{\struct{N}}(\struct{X}\cap\struct{Y})$, respectively, giving the
displayed inequality.
\end{proof}

For finite $\struct{X}$ and arbitrary $\struct{Z}$, define the \emph{relative dimension of $\struct{X}$ over $\struct{Z}$} by
\[
  d_{\struct{N}}(\struct{X}/\struct{Z})
  \coloneqq
  \inf\{d_{\struct{N}}(\struct{X}\cup\struct{Z}_0)-d_{\struct{N}}(\struct{Z}_0):\struct{Z}_0\subseteq\struct{Z}\text{ finite}\}.
\]

\begin{enumerate}
\renewcommand{\theenumi}{D\arabic{enumi}}
\renewcommand{\labelenumi}{(\theenumi)}
\setcounter{enumi}{1}

\item\label{property:appendix-base-enlargement-rel-dimension} \textbf{Enlargement of base does not increase relative dimension.}
Relative dimension is non-negative and does not increase when the base is enlarged: for arbitrary $\struct{Z}\subseteq\struct{W}\subseteq\struct{N}$,
\[
  0\leq d_{\struct{N}}(\struct{X}/\struct{W})\leq d_{\struct{N}}(\struct{X}/\struct{Z}).
\]
For a finite base $\struct{B}$, its definition simplifies to
\[
  d_{\struct{N}}(\struct{X}/\struct{B})
  =d_{\struct{N}}(\struct{X}\cup\struct{B})-d_{\struct{N}}(\struct{B}).
\]
\end{enumerate}
\begin{proof}
If $\struct{C}\subseteq\struct{B}$ are finite, submodularity applied to $\struct{X}\cup\struct{C}$ and $\struct{B}$, together with monotonicity and $\struct{C}\subseteq(\struct{X}\cup\struct{C})\cap\struct{B}$, gives
\[
  d_{\struct{N}}(\struct{X}\cup\struct{B})
  -d_{\struct{N}}(\struct{B})
  \leq d_{\struct{N}}(\struct{X}\cup\struct{C})
  -d_{\struct{N}}(\struct{C}).
\]
These differences are nonnegative, again by monotonicity.
Thus, for a finite base $\struct{B}$, the infimum defining $d_{\struct{N}}(\struct{X}/\struct{B})$ is attained at $\struct{B}$ itself.
For arbitrary bases, enlarging the base enlarges the collection of finite subsets over which the infimum is taken, so relative dimension does not increase.
\end{proof}

\begin{enumerate}[(D3)]
\item \textbf{Invariance under closure.}
Taking the closure of either the finite argument or the base does
not change the relative dimension: For every $\struct{Z}\subseteq\struct{N}$
\[
  d_{\struct{N}}(\cl_{\struct{N}}(\struct{X})/\struct{Z})
  =d_{\struct{N}}(\struct{X}/\struct{Z})
  =d_{\struct{N}}(\struct{X}/\cl_{\struct{N}}(\struct{Z})).
\]
\end{enumerate}
\begin{proof}
We first prove the finite identity
\begin{equation}\label{eq:appendix-closure-invariance-finite}
  d_{\struct{N}}(\struct{U}\cup\cl_{\struct{N}}(\struct{V}))
  =d_{\struct{N}}(\struct{U}\cup\struct{V}).
\end{equation}
Submodularity applied to $\struct{U}\cup\struct{V}$ and $\cl_{\struct{N}}(\struct{V})$ gives
\[
  d_{\struct{N}}(\struct{U}\cup\cl_{\struct{N}}(\struct{V}))
  +d_{\struct{N}}((\struct{U}\cup\struct{V})\cap\cl_{\struct{N}}(\struct{V}))
  \leq d_{\struct{N}}(\struct{U}\cup\struct{V})+d_{\struct{N}}(\struct{V}).
\]
The intersection contains $\struct{V}$, so monotonicity gives $d_{\struct{N}}(\struct{U}\cup\cl_{\struct{N}}(\struct{V}))\leq d_{\struct{N}}(\struct{U}\cup\struct{V})$. The reverse inequality follows from $\struct{U}\cup\struct{V}\subseteq\struct{U}\cup\cl_{\struct{N}}(\struct{V})$. Applying this identity with $\struct{U}$ a finite subset of $\struct{Z}$ and $\struct{V}=\struct{X}$ shows $d_{\struct{N}}(\cl_{\struct{N}}(\struct{X})/\struct{Z})=d_{\struct{N}}(\struct{X}/\struct{Z})$.

It remains to prove invariance under closure of the base.
Since $\struct{Z}\subseteq\cl_{\struct{N}}(\struct{Z})$, property~\eqref{property:appendix-base-enlargement-rel-dimension} gives
$d_{\struct{N}}(\struct{X}/\cl_{\struct{N}}(\struct{Z}))\leq d_{\struct{N}}(\struct{X}/\struct{Z})$.
For the reverse inequality, let $\struct{W}\subseteq\cl_{\struct{N}}(\struct{Z})$ be finite.
Finite character gives a finite $\struct{Z}_0\subseteq\struct{Z}$ with
$\struct{W}\subseteq\cl_{\struct{N}}(\struct{Z}_0)$.
The inclusions
\[
  \struct{Z}_0\subseteq\struct{W}\cup\struct{Z}_0\subseteq\cl_{\struct{N}}(\struct{Z}_0)
\]
and \eqref{eq:appendix-closure-invariance-finite} imply
\[
  d_{\struct{N}}(\struct{W}\cup\struct{Z}_0)=d_{\struct{N}}(\struct{Z}_0),
  \qquad
  d_{\struct{N}}(\struct{X}\cup\struct{W}\cup\struct{Z}_0)=d_{\struct{N}}(\struct{X}\cup\struct{Z}_0).
\]
Consequently,
\[
  d_{\struct{N}}(\struct{X}/\struct{W})
  \geq d_{\struct{N}}(\struct{X}/(\struct{W}\cup\struct{Z}_0))
  =d_{\struct{N}}(\struct{X}/\struct{Z}_0)
  \geq d_{\struct{N}}(\struct{X}/\struct{Z}).
\]
Taking the infimum over all finite $\struct{W}\subseteq\cl_{\struct{N}}(\struct{Z})$ proves
the reverse inequality.
\end{proof}
\begin{enumerate}[(D4)]
\item \textbf{The chain rule.}
For every $\struct{Z}\subseteq\struct{N}$,
\begin{equation}\label{eq:appendix-chain-rule}
  d_{\struct{N}}((\struct{X}\cup\struct{Y})/\struct{Z})
  =d_{\struct{N}}(\struct{X}/(\struct{Y}\cup\struct{Z}))
  +d_{\struct{N}}(\struct{Y}/\struct{Z}).
\end{equation}
\end{enumerate}
\begin{proof}
If the base $\struct{B}$ is finite, cancellation gives
\[
  d_{\struct{N}}((\struct{X}\cup\struct{Y})/\struct{B})
  =d_{\struct{N}}(\struct{X}/(\struct{Y}\cup\struct{B}))
  +d_{\struct{N}}(\struct{Y}/\struct{B}).
\]
For an arbitrary base $\struct{Z}$, let $\struct{B}$ range over its finite subsets. Since $\struct{Y}$ is finite, every finite subset of $\struct{Y}\cup\struct{Z}$ is contained
in some $\struct{Y}\cup\struct{B}$. By~property~\eqref{property:appendix-base-enlargement-rel-dimension}, it therefore suffices to use bases of
the form $\struct{Y}\cup\struct{B}$ when computing $d_{\struct{N}}(\struct{X}/(\struct{Y}\cup\struct{Z}))$.

Both terms $d_{\struct{N}}(\struct{X}/(\struct{Y}\cup\struct{B}))$ and $d_{\struct{N}}(\struct{Y}/\struct{B})$ are non-increasing
as $\struct{B}$ grows. Given $\varepsilon>0$, choose finite subsets of $\struct{Z}$
approximating their respective infima within $\varepsilon/2$.
Replacing these subsets by their union approximates both infima
simultaneously. Hence the infimum of the sum is the sum of the
infima, proving \eqref{eq:appendix-chain-rule}.
\end{proof}

\paragraph*{The extension property in arbitrary models.}
The final type count takes place in a saturated elementary
extension of $\struct{M}_\alpha$. We therefore show that every model of
$T_\alpha$ satisfies the finite extension property, using the
first-order definability of finite closedness. Let
\[
  b(k)\coloneqq
  \left\lfloor F_\alpha^{-1}(\alpha k)\right\rfloor.
\]
If a $k$-element set $\struct{A}$ is not closed in a graph whose finite subgraphs lie
in $\mathcal C_\alpha$, then some finite $\struct{B}\supseteq\struct{A}$ has
$\partial(\struct{B})<\partial(\struct{A})$. Applying the closure construction inside $\struct{B}$ and
the estimate \eqref{eq:closure-bound} gives such a witness with at most
$b(k)$ vertices. Thus, closedness of a fixed finite tuple is first-order
definable: one excludes the finitely many diagrams of witnesses of size at
most $b(k)$.

For every finite strong extension $\struct{A}\leq_\alpha\struct{B}$, the limit graph
$\struct{M}_\alpha$ satisfies the first-order sentence saying that every closed copy
of $\struct{A}$ extends to a closed copy of $\struct{B}$. These sentences therefore belong to
$T_\alpha\coloneqq\operatorname{Th}(\struct{M}_\alpha)$. It follows that every
$\struct{N}\models T_\alpha$ has the following two properties:
\begin{enumerate}
\renewcommand{\theenumi}{F\arabic{enumi}}
\renewcommand{\labelenumi}{(\theenumi)}

\item\label{property:appendix-finite-substructures-in-C-alpha} every finite induced subgraph of $\struct{N}$ belongs to $\mathcal C_\alpha$;
\item\label{property:appendix-finite-extension-property} if $\struct{A}\leq_\alpha\struct{N}$ is finite and $\struct{A}\leq_\alpha\struct{B}$ with
$\struct{B}\in\mathcal C_\alpha$, then $\struct{N}$ contains over $\struct{A}$ a closed copy of $\struct{B}$.
\end{enumerate}

\paragraph*{Closed diagrams and free amalgamation.}
We first show that isomorphisms between closed subgraphs are elementary. We then prove that dimension independence yields
a closed free amalgam.

\begin{lemma}[Closed diagrams]\label[lemma]{lem:closed-diagrams}
Let $\struct{N}_1,\struct{N}_2\models T_\alpha$. Every isomorphism between closed induced
subgraphs of $\struct{N}_1$ and $\struct{N}_2$ is elementary.
\end{lemma}

\begin{proof}
The isomorphisms between finite closed subgraphs form a back-and-forth
system. Indeed, let $h\colon\struct{E}_1\to\struct{E}_2$ be such an isomorphism and let
$c\in\struct{N}_1$. Then $\struct{D}\coloneqq\cl_{\struct{N}_1}(\struct{E}_1c)$ is a finite closed strong
extension of $\struct{E}_1$. Property~\eqref{property:appendix-finite-extension-property} in $\struct{N}_2$ gives a closed copy of $\struct{D}$ over
$\struct{E}_2$, extending $h$; the converse direction is symmetric. Hence every
isomorphism between finite closed subgraphs is elementary.

Now let $h\colon\struct{A}_1\to\struct{A}_2$ be an isomorphism between arbitrary
closed subgraphs. Every finite tuple from $\struct{A}_1$ is contained in
a finite subgraph $\struct{E}_1\subseteq\struct{A}_1$ closed in $\struct{N}_1$.
Put $\struct{D}\coloneqq\cl_{\struct{N}_2}(h(\struct{E}_1))$. Since $\struct{A}_2$ is closed,
monotonicity gives $\struct{D}\subseteq\struct{A}_2$. By the definition of closure,
\[
  \partial(h^{-1}(\struct{D}))=\partial(\struct{D})
  \leq\partial(h(\struct{E}_1))=\partial(\struct{E}_1).
\]
Since $\struct{E}_1$ is closed in $\struct{N}_1$, no proper finite extension of
$\struct{E}_1$ can have predimension at most $\partial(\struct{E}_1)$.
Thus, $h^{-1}(\struct{D})=\struct{E}_1$, so $\struct{D}=h(\struct{E}_1)$.
The restriction $h|_{\struct{E}_1}$ is therefore elementary by the preceding
paragraph. Since every finite tuple from $\struct{A}_1$ is contained in
such an $\struct{E}_1$, it follows that $h$ is elementary.
\end{proof}
For closed substructures $\struct{A},\struct{B},\struct{C}$ of $\struct{N}$ with $\struct{C}\subseteq\struct{A}\cap\struct{B}$, write
$\struct{A}\mathrel{\mathop{\perp}\limits^d_{\struct{C}}}\struct{B}$ if
\begin{itemize}
\item
$\cl_{\struct{N}}(\struct{A}\cup\struct{C})\cap\cl_{\struct{N}}(\struct{B}\cup\struct{C})=\cl_{\struct{N}}(\struct{C})$ and
\item
for every finite set $\struct{U}\subseteq\struct{A}$, we have $d_{\struct{N}}(\struct{U}/\struct{B})=d_{\struct{N}}(\struct{U}/\struct{C})$.
\end{itemize}
The next lemma is the
separation step in Wagner's argument.

\begin{lemma}[Dimension separation]\label[lemma]{lem:dimension-separation}
If $\struct{A},\struct{B}\leq_\alpha\struct{N}$ and
$\struct{A}\mathrel{\mathop{\perp}\limits^d_{\struct{A}\cap\struct{B}}}\struct{B}$, then $\struct{A}\cup\struct{B}$ is their
free amalgam over $\struct{A}\cap\struct{B}$ and is closed in $\struct{N}$.
\end{lemma}

\begin{proof}
First suppose that $\struct{A}$ and $\struct{B}$ are finite, and put $\struct{C}\coloneqq\struct{A}\cap\struct{B}$.
Since all three sets are closed, dimension independence gives
\[
  d_{\struct{N}}(\struct{A}\cup\struct{B})+d_{\struct{N}}(\struct{C})
  =d_{\struct{N}}(\struct{A})+d_{\struct{N}}(\struct{B})
  =\partial(\struct{A})+\partial(\struct{B}).
\]
Direct edge counting gives
\begin{equation}\label{eq:appendix-separation}
  \partial(\struct{A})+\partial(\struct{B})-\partial(\struct{A}\cup\struct{B})-\partial(\struct{C})
  =\alpha^{-1}e(\struct{A}\setminus\struct{B},\struct{B}\setminus\struct{A}).
\end{equation}
Here $e(\struct{A}\setminus\struct{B}, \struct{B}\setminus\struct{A})$ counts the edges between $A\setminus B$ and $B\setminus A$. If the amalgam were not free, the right-hand side would be at least $1/\alpha$. Since $d_{\struct{N}} (\struct{A}\cup\struct{B})\leq\partial(\struct{A}\cup\struct{B})$, this would contradict
the displayed dimension equality. Thus, the amalgam is free. Equality in
\eqref{eq:appendix-separation} now gives
$d_{\struct{N}}(\struct{A}\cup\struct{B})=\partial(\struct{A}\cup\struct{B})$, so $\struct{A}\cup\struct{B}$ is closed.

Now let $\struct{A}$ and $\struct{B}$ be arbitrary. We first exclude a cross-edge
$ab$ with $a\in\struct{A}\setminus\struct{B}$ and $b\in\struct{B}\setminus\struct{A}$. Put
$\struct{C}\coloneqq\struct{A}\cap\struct{B}$ and $\gamma\coloneqq1/\alpha$. Choose a finite
$\struct{C}_0\subseteq\struct{C}$ such that
\[
  d_{\struct{N}}(a/\struct{C}_0)<d_{\struct{N}}(a/\struct{C})+\gamma/3,
\]
and put $\struct{A}_0\coloneqq\cl_{\struct{N}}(\{a\}\cup\struct{C}_0)$ and $\struct{C}_1\coloneqq\struct{A}_0\cap\struct{C}$.
Choose a finite closed $\struct{B}_0\subseteq\struct{B}$ containing $\struct{C}_1\cup\{b\}$, and set
$\struct{C}_2\coloneqq\struct{A}_0\cap\struct{B}_0$. The sets $\struct{A}_0,\struct{B}_0,\struct{C}_2$ are closed, and the
cross-edge $ab$ gives
\[
  \partial(\struct{A}_0\cup\struct{B}_0)-\partial(\struct{B}_0)
  \leq\partial(\struct{A}_0)-\partial(\struct{C}_2)-\gamma
\]
by \eqref{eq:appendix-separation}. Therefore
\[
  d_{\struct{N}}(\struct{A}_0/\struct{B}_0)
  \leq\partial(\struct{A}_0\cup\struct{B}_0)-\partial(\struct{B}_0)
  \leq d_{\struct{N}}(\struct{A}_0/\struct{C}_2)-\gamma.
\]
On the other hand, independence and monotonicity give
\[
  d_{\struct{N}}(\struct{A}_0/\struct{C})\leq d_{\struct{N}}(\struct{A}_0/\struct{B}_0),
  \qquad
  d_{\struct{N}}(\struct{A}_0/\struct{C}_2)\leq d_{\struct{N}}(\struct{A}_0/\struct{C}_1)
  <d_{\struct{N}}(\struct{A}_0/\struct{C})+\gamma/3,
\]
a contradiction. Hence the amalgam is free.

Finally, suppose for a contradiction that $\struct{U}\coloneqq\struct{A}\cup\struct{B}$ is not
closed. Choose a finite $\struct{S}\subseteq\struct{U}$ such that
$\cl_{\struct{N}}(\struct{S})$ is not contained in $\struct{U}$, and put
$\struct{S}_0\coloneqq\cl_{\struct{N}}(\struct{S})\cap\struct{U}$. Since $\struct{S}\subseteq\struct{S}_0\subsetneq\cl_{\struct{N}}(\struct{S})$ and $\cl_{\struct{N}}(\struct{S})$ is a
minimiser for $\struct{S}$, we have $\partial(\cl_{\struct{N}}(\struct{S}))\leq\partial(\struct{S}_0)$. Equality is
impossible for the proper extension $\struct{S}_0\subsetneq\cl_{\struct{N}}(\struct{S})$ because $\alpha$ is
irrational. Thus
\[
  \varepsilon\coloneqq\partial(\struct{S}_0)-\partial(\cl_{\struct{N}}(\struct{S}))>0.
\]
This drop persists under finite enlargement inside $\struct{U}$. Indeed, if
$\struct{U}_0\subseteq\struct{U}$ is finite and contains $\struct{S}_0$, then
$\cl_{\struct{N}}(\struct{S})\cap\struct{U}_0=\struct{S}_0$, so submodularity gives
\begin{equation}\label{eq:appendix-fixed-drop}
  d_{\struct{N}}(\struct{U}_0)
  \leq\partial(\cl_{\struct{N}}(\struct{S})\cup\struct{U}_0)
  \leq\partial(\struct{U}_0)-\varepsilon.
\end{equation}

Choose finite closed $\struct{A}_1\subseteq\struct{A}$ and $\struct{B}_1\subseteq\struct{B}$ containing
$\struct{S}_0\cap\struct{A}$ and $\struct{S}_0\cap\struct{B}$, respectively. Put $\struct{C}\coloneqq\struct{A}\cap\struct{B}$.
By the definition of relative dimension, choose a finite
$\struct{C}_0\subseteq\struct{C}$, containing $\struct{A}_1\cap\struct{B}_1$, such that
\[
  d_{\struct{N}}(\struct{A}_1/\struct{C}_0)<d_{\struct{N}}(\struct{A}_1/\struct{C})+\varepsilon/2.
\]
Set
\[
  \struct{A}_2\coloneqq\cl_{\struct{N}}(\struct{A}_1\cup\struct{C}_0),
  \qquad
  \struct{B}_2\coloneqq\cl_{\struct{N}}(\struct{B}_1\cup\struct{C}_0),
  \qquad
  \struct{C}_2\coloneqq\struct{A}_2\cap\struct{B}_2.
\]
Because $\struct{A}$ and $\struct{B}$ are closed, $\struct{A}_2\subseteq\struct{A}$ and $\struct{B}_2\subseteq\struct{B}$.
Thus, $\struct{U}_0\coloneqq\struct{A}_2\cup\struct{B}_2$ is a finite subset of $\struct{U}$ containing
$\struct{S}_0$, and \eqref{eq:appendix-fixed-drop} applies. Since $\struct{B}_2$ and $\struct{C}_2$
are closed, the chain rule and submodularity yield
\begin{align*}
  d_{\struct{N}}(\struct{A}_2/\struct{B}_2)
  &=d_{\struct{N}}(\struct{U}_0)-\partial(\struct{B}_2)\\
  &\leq\partial(\struct{U}_0)-\partial(\struct{B}_2)-\varepsilon\\
  &\leq\partial(\struct{A}_2)-\partial(\struct{C}_2)-\varepsilon\\
  &=d_{\struct{N}}(\struct{A}_2/\struct{C}_2)-\varepsilon.
\end{align*}
As $\struct{C}_0\subseteq\struct{C}_2$, monotonicity and closure invariance give
\[
  d_{\struct{N}}(\struct{A}_2/\struct{C}_2)
  \leq d_{\struct{N}}(\struct{A}_2/\struct{C}_0)
  =d_{\struct{N}}(\struct{A}_1/\struct{C}_0),
  \qquad
  d_{\struct{N}}(\struct{A}_2/\struct{C})=d_{\struct{N}}(\struct{A}_1/\struct{C}).
\]
Consequently,
\[
  d_{\struct{N}}(\struct{A}_2/\struct{B})
  \leq d_{\struct{N}}(\struct{A}_2/\struct{B}_2)
  <d_{\struct{N}}(\struct{A}_2/\struct{C})-\varepsilon/2,
\]
contradicting $\struct{A}\mathrel{\mathop{\perp}\limits^d_{\struct{C}}}\struct{B}$. Hence $\struct{A}\cup\struct{B}$
is closed.
\end{proof}

\paragraph*{Counting types.}
We now bound the number of types over a parameter set of cardinality
$2^{\aleph_0}$, using the preceding two lemmata to prove stability.

\begin{theorem}\label[theorem]{thm:appendix-stability}
The theory $T_\alpha=\operatorname{Th}(\struct{M}_\alpha)$ is stable.
\end{theorem}

\begin{proof}
Let $\lambda\coloneqq2^{\aleph_0}$. In ZFC, $\lambda^{\aleph_0}=(2^{\aleph_0})^{\aleph_0}=2^{\aleph_0\cdot\aleph_0}=\lambda$. We show that $T_\alpha$ is
$\lambda$-stable. It suffices to consider parameter sets of cardinality $\lambda$: smaller sets can be enlarged in an elementary extension, and their types extend to types over the larger set. Let $\struct{M}_0\models T_\alpha$, let $\struct{P}\subseteq\struct{M}_0$ have
cardinality $\lambda$, and take a $\lambda^+$-saturated elementary extension
$\struct{N}\succeq\struct{M}_0$. Put $\struct{X}\coloneqq\cl_{\struct{N}}(\struct{P})$. Since closure is finitary and
finite sets have finite closure, $|\struct{X}|=\lambda$.

Every finite type over $\struct{P}$ is realised in $\struct{N}$. Fix a finite tuple $\bar a$ (we use $\bar{a}$ for both the tuple and the set of its entries)
from $\struct{N}$. By the definition of relative dimension, there are finite
$\struct{Y}_n\subseteq\struct{X}$ such that
\[
  d_{\struct{N}}(\bar a/\struct{Y}_n)<d_{\struct{N}}(\bar a/\struct{X})+2^{-n}.
\]
Let $\struct{X}_0\coloneqq\bigcup_n\struct{Y}_n$. Then $\struct{X}_0$ is countable and
$d_{\struct{N}}(\bar a/\struct{X}_0)=d_{\struct{N}}(\bar a/\struct{X})$. Put
\[
  \struct{A}\coloneqq\cl_{\struct{N}}(\bar a\cup\struct{X}_0),
  \qquad
  \struct{X}_1\coloneqq\struct{A}\cap\struct{X}.
\]
Both $\struct{A}$ and $\struct{X}_1$ are countable and closed, and $\struct{X}_0\subseteq\struct{X}_1$.
Monotonicity and closure invariance give
\[
  d_{\struct{N}}(\bar a/\struct{X})
  \leq d_{\struct{N}}(\bar a/\struct{X}_1)
  \leq d_{\struct{N}}(\bar a/\struct{X}_0)
  =d_{\struct{N}}(\bar a/\struct{X}),
\]
We claim that the same equality holds for every finite tuple $\bar b$ from
$\struct{A}$. By finitarity of closure, choose a finite $\struct{X}'_0\subseteq\struct{X}_0$ such
that $\bar b\subseteq\cl_{\struct{N}}(\bar a\struct{X}'_0)$. Since $\struct{X}'_0\subseteq\struct{X}_1\subseteq\struct{X}$, closure invariance and
monotonicity in the finite argument give
\[
  d_{\struct{N}}((\bar a\cup\bar b)/\struct{X})=d_{\struct{N}}(\bar a/\struct{X}),
  \qquad
  d_{\struct{N}}((\bar a\cup\bar b)/\struct{X}_1)=d_{\struct{N}}(\bar a/\struct{X}_1).
\]
The right-hand sides are equal. Applying the chain rule to the left-hand sides, we obtain
\[
  d_{\struct{N}}(\bar b/\struct{X})+d_{\struct{N}}(\bar a/(\bar b\cup\struct{X}))
  =
  d_{\struct{N}}(\bar b/\struct{X}_1)+d_{\struct{N}}(\bar a/(\bar b\cup\struct{X}_1)).
\]
Because $\struct{X}_1\subseteq\struct{X}$, each summand on the left is at most the corresponding summand on the right. Equality of their sums forces equality of the corresponding summands;
in particular,
\[
  d_{\struct{N}}(\bar b/\struct{X})=d_{\struct{N}}(\bar b/\struct{X}_1).
\]
Moreover, $\struct{A}$ and $\struct{X}$ are closed, $\struct{A}\cap\struct{X}=\struct{X}_1$, and $\struct{X}_1$ is closed.
Hence
\[
  \cl_{\struct{N}}(\struct{A}\cup\struct{X}_1)\cap\cl_{\struct{N}}(\struct{X}\cup\struct{X}_1)
  =\struct{A}\cap\struct{X}
  =\struct{X}_1
  =\cl_{\struct{N}}(\struct{X}_1).
\]
Thus, both conditions in the definition of independence hold, so
$\struct{A}\mathrel{\mathop{\perp}\limits^d_{\struct{X}_1}}\struct{X}$. By
\Cref{lem:dimension-separation}, $\struct{A}\cup\struct{X}$ is the free amalgam of $\struct{A}$ and
$\struct{X}$ over $\struct{X}_1$ and is closed in $\struct{N}$.

We now count the possibilities for $\operatorname{tp}(\bar a/\struct{X})$. There are
at most
\[
  |\struct{X}|^{\aleph_0}=\lambda
\]
choices for the countable set $\struct{X}_1\subseteq\struct{X}$, and at most
$2^{\aleph_0}=\lambda$ possibilities for the countable diagram of $\struct{A}$ with
the distinguished tuple $\bar a$ and its identified subgraph $\struct{X}_1$. Since
$\struct{A}\cup\struct{X}$ is a free amalgam, these data determine its complete diagram.
By \Cref{lem:closed-diagrams}, that diagram determines the complete type of
$\struct{A}\cup\struct{X}$, and hence the type of $\bar a$ over $\struct{X}$. Therefore there are at
most $\lambda$ finite types over $\struct{X}$, and consequently at most $\lambda$
finite types over $\struct{P}$. Thus, $T_\alpha$ is $\lambda$-stable and hence stable.
\end{proof}

\section{The open-dense parameter argument}\label{app:evans-density}

We reconstruct the Baire-category step from \cite[Example~3.22]{EvansBonn}. Recall that
\[
  \eta(\alpha,N)=
  \min\left\{\alpha-\frac ab:
  a,b\in\NN,\ 1\leq b\leq N,\ \frac ab<\alpha\right\}.
\]
The strict inequality differs from Evans' convention only at rational
parameters.  For the finite partial sum used below, our interval contains no
rational $a/b$ with $b$ within the relevant denominator bound.  Thus, no
parameter in the interval is equal to an admissible rational approximation,
and the proof applies with the present convention.

\begin{proof}[Proof of \Cref{lem:comeager-continuum}]
Construct closed rational intervals $J_s$ of positive length indexed by binary strings
$s\in2^{<\omega}$.  Start with $J_\varnothing\subseteq I$.  Having chosen
$J_s$, use openness and density of $V_{|s|}$ to choose two disjoint closed rational
subintervals
\[
  J_{s0},J_{s1}\subseteq
  \operatorname{int}(J_s)\cap V_{|s|}
\]
of diameter at most $2^{-|s|-1}$.  For every infinite binary sequence
$x\in2^\omega$, the nested intervals $J_{x\upharpoonright n}$ have a unique
common point.  Distinct sequences give distinct points because sibling
intervals are disjoint, and every resulting point lies in every $V_m$.
The resulting map from $2^\omega$ to the intersection is continuous and
injective.  Since $2^\omega$ is compact and $\mathbb R$ is Hausdorff, it is a
homeomorphism onto its image.  The reverse cardinality bound holds because
the intersection is a subset of $\mathbb R$.
\end{proof}

\begin{lemma}[Denominator bound]\label[lemma]{lem:rational-denominator-bound}
If
\[
  \frac ab<\frac cd,
  \qquad bc-ad=1,
\]
then every rational $p/q$ strictly between them has $q\geq b+d$.
\end{lemma}

\begin{proof}
\[
  \frac1{bd}
  =\frac cd-\frac ab
  =\left(\frac pq-\frac ab\right)
   +\left(\frac cd-\frac pq\right)
  \geq\frac1{bq}+\frac1{dq}
  =\frac{b+d}{bdq}.
\]
Thus, $q\geq b+d$.
\end{proof}

\begin{proposition}[Open-dense parameter intervals]
\label[proposition]{prop:appendix-evans-density}
For every nonempty open interval $I\subseteq(1,\infty)$ and every
$m\in\NN$, there are $K\in\NN$ and a nonempty rational open interval $J$
with $\overline J\subseteq I$ such that
\[
  \sum_{N=1}^K\eta(\alpha,N)>m
  \qquad\text{for every }\alpha\in J.
\]
Consequently, for each $m$, the union of all such intervals $J$ is open and
dense in $(1,\infty)$.
\end{proposition}

\begin{proof}
Choose a reduced fraction $c/d\in I$ with $d>0$, and choose $\varepsilon>0$ such that
$(c/d-\varepsilon,c/d]\subseteq I$. There are integers $a_0,b_0$ with
$b_0c-a_0d=1$. For $t\in\NN$, put
\[
  a_t\coloneqq a_0+tc,
  \qquad
  b_t\coloneqq b_0+td.
\]
Then $b_tc-a_td=1$. Choose $t$ so large that
\[
  b_t>\frac1{d\varepsilon}.
\]
It follows that
\[
  0<\frac cd-\frac{a_t}{b_t}
  =\frac1{db_t}<\varepsilon.
\]
Thus, $a_t/b_t\in I$. Rename $a_t,b_t$ as $a,b$; then
\[
  \frac ab\in I,
  \qquad
  \frac ab<\frac cd.
\]
For $k\geq0$, put
\[
  a_k\coloneqq a+kc,
  \qquad
  b_k\coloneqq b+kd,
  \qquad
  r_k\coloneqq\frac{a_k}{b_k}.
\]
Then $b_kc-a_kd=1$, so $r_k<c/d$, the sequence $(r_k)$ increases to
$c/d$, and by \Cref{lem:rational-denominator-bound} every rational strictly between $r_k$ and $c/d$ has denominator
at least $b_k+d=b_{k+1}.$
Choose $L$ so large that
\begin{equation}\label{eq:appendix-harmonic-blocks}
  \frac12\sum_{k=0}^L\frac1{b_k}>m.
\end{equation}
Choose a nonempty rational open interval $J$ whose closure is contained in
\[
  \left(\frac cd-\frac1{2db_{L+1}},\frac cd\right)
  \subseteq I.
\]
The left endpoint of the displayed interval is greater than $r_{L+1}$,
because
\[
  \frac cd-r_{L+1}=\frac1{db_{L+1}}.
\]
Thus, every $\alpha\in J$ lies above $r_{L+1}$.

Fix such an $\alpha$. If $0\leq k\leq L$ and
$b_k\leq N<b_{k+1}$, \Cref{lem:rational-denominator-bound} shows that
$r_k$ is the greatest rational below $\alpha$ with denominator at most $N$.
Hence
\[
  \eta(\alpha,N)=\alpha-r_k.
\]
Moreover,
\[
  \alpha-r_k
  =\left(\frac cd-r_k\right)-\left(\frac cd-\alpha\right)
  >\frac1{db_k}-\frac1{2db_{L+1}}
  \geq\frac1{2db_k}.
\]
There are exactly $b_{k+1}-b_k=d$ integers $N$ in this block. Therefore,
with $K\coloneqq b_{L+1}-1$,
\[
  \sum_{N=1}^K\eta(\alpha,N)
  \geq
  \sum_{k=0}^L\sum_{N=b_k}^{b_{k+1}-1}\eta(\alpha,N)
  >\frac12\sum_{k=0}^L\frac1{b_k}
  >m
\]
by \eqref{eq:appendix-harmonic-blocks}. The same $K$ works for every
$\alpha\in J$. Since $I$ was arbitrarily chosen, the union of all such $J$ is dense;
it is open by definition.
\end{proof}

\section{High girth and density}\label{app:high-girth-density}

\begin{proof}[Proof of \Cref{lem:high-girth-density}]
Choose integers $k>2q+1$ and $g'>\max\{g,L\}$. By Erd\H{o}s's theorem there is a finite graph $\struct{H}_0$ with chromatic number at least $k$ and girth at least $g'$. Let $\struct{H}\subseteq\struct{H}_0$ be minimal under vertex deletion subject to $\chi(\struct{H})\geq k$. Then every vertex of $\struct{H}$ has degree at least $k-1$: otherwise a $(k-1)$-colouring of $\struct{H}-v$ would extend to $v$. Hence
\[
  2e(\struct{H})\geq(k-1)|\struct{H}|>2q|\struct{H}|.
\]
Moreover, $\struct{H}$ contains a cycle because $k\geq3$, and every cycle has length at least $g'>\max\{g,L\}$. Thus, $|\struct{H}|\geq L$, the girth is greater than $g$, and $e(\struct{H})>q|\struct{H}|$.
\end{proof}

\section{CSP and effective parameter calculations}
\label{app:effective-details}

\begin{proof}[Proof of \Cref{lem:complete-age}]
Every pair of distinct vertices of
$\struct{X}_{\struct{G}}$ carries either an
$E$-constraint or an $N$-constraint.
Both relations are irreflexive in $\struct{B}_\alpha$,
so any homomorphism is injective.
It preserves edges and non-edges and hence gives
an induced embedding of $\struct{G}$ into
$\struct{M}_\alpha$.
Conversely, every such induced embedding preserves
both relations and gives a homomorphism from
$\struct{X}_{\struct{G}}$ to $\struct{B}_\alpha$.
Finally, $\Age(\struct{M}_\alpha)=\mathcal C_\alpha$.
\end{proof}

\begin{proof}[Proof of \Cref{lem:quotient-completion}]
If $h\colon\struct{X}\to\struct{B}_\alpha$, its kernel $\sim$ is consistent by irreflexivity and disjointness of $E$ and $N$ in $\struct{B}_\alpha$. The graph induced by $E$ on the image of $h$, transferred to the quotient classes, belongs to $\mathcal C_\alpha$. Deleting its edges that are not required by $E$-constraints in $\struct{X}$ gives $\struct{G}_\sim$, which still belongs to $\mathcal C_\alpha$.

Conversely, if $\sim$ is consistent and $\struct{G}_\sim\in\mathcal C_\alpha$, choose an induced embedding $\struct{G}_\sim\hookrightarrow\struct{M}_\alpha$. Its composition with the quotient map preserves every $E$- and $N$-constraint and hence gives a homomorphism to $\struct{B}_\alpha$.
\end{proof}

\begin{proof}[Proof of \Cref{lem:finite-stratification}]
Only approximation denominators $1\leq d\leq r$ occur. Fix
$\alpha\in[a,b]$. Among the fractions with denominator exactly $d$, the
largest one strictly below $\alpha$ is
$\frac{\lceil d\alpha\rceil-1}{d}$.
Indeed, its numerator is the largest integer strictly smaller than
$d\alpha$. Consequently, if $c/d$ is the best lower approximation to
$\alpha$ among fractions with denominator at most $N$, for some
$N\in\{d,\ldots,r\}$, then we have $c=\lceil d\alpha\rceil-1$.
It follows that
\(
  da-1
  \leq d\alpha-1
  \leq c
  <d\alpha
  \leq db.
\)

Thus, for each $d\leq r$, only finitely many integer numerators $c$ can
occur. Let $Q$ contain $a$, $b$, and all the resulting rationals $c/d$ that
lie in $[a,b]$. Then $Q\subseteq[a,b]\cap\QQ$. On one component of the
complement, the order relation between the
parameter $\alpha$ and every relevant rational is fixed. Hence, for each
$N\leq r$, the greatest relevant rational below $\alpha$ is a fixed number
$q_N$, and
\[
  \eta(\alpha,N)=\alpha-q_N.
\]
It follows that
$F_\alpha(m)=m\alpha-\sum_{N=1}^m q_N$ on the component. All rationals and
all coefficients arise from finite searches, so the decomposition is
effective. The strict convention in the definition of $\eta$ affects the
values at points of $Q$, but not on the open components used in the interval
construction.
\end{proof}

\section{A parameter for the polynomial-hierarchy bound}
\label{app:explicit-parameter}
The following parameter comes from Pantano's construction~\cite[Appendix~B]{Pantano1995}, translated to our reciprocal
parameterisation.  We include the comparison analysis because the large gaps
between successive decimal positions do not prevent polynomial-time
comparison at the relevant input scale.

\begin{proposition}[Pantano's explicit parameter]
\label[proposition]{prop:pantano-explicit-parameter}
Let
\[
  d_1=1,
  \qquad
  d_{m+1}=d_m+10^{2d_m}+1,
\]
and define
\[
  \beta\coloneqq\sum_{m=1}^{\infty}10^{-d_m},
  \qquad
  \alpha\coloneqq\frac1\beta.
\]
Then $\alpha$ is an irrational member of $\mathcal A$ with
$9<\alpha<10$.  Moreover, given $N\geq1$ and integers $a,b$ satisfying
\[
  0\leq a\leq10N,
  \qquad
  1\leq b\leq N,
\]
one can decide whether $a/b<\alpha$ in time polynomial in $N$.
\end{proposition}

\begin{proof}
The first term of $\beta$ is $1/10$, and the remaining nonzero decimal digits occur after the second decimal place.  Hence $1/10<\beta<1/9$, so $9<\alpha<10$.  The gaps $d_{m+1}-d_m$ tend to infinity.  Thus, the decimal expansion of $\beta$ is not eventually periodic: for any proposed period, a sufficiently isolated digit $1$ would force another digit $1$ before the next $d_{m+1}$. Consequently, $\beta$ and $\alpha$ are irrational.

Pantano uses the approximation function
\[
  \varepsilon(\beta,N)
  =\min\left\{1-\frac pq\beta>0:
      p,q\in\NN,\ 1\leq q\leq N\right\}.
\]
Since $\alpha=1/\beta$,
\[
  1-\frac pq\beta
  =\beta\left(\alpha-\frac pq\right),
\]
and therefore
\[
  \varepsilon(\beta,N)=\beta\eta(\alpha,N),
  \qquad
  \sum_{j=1}^N\varepsilon(\beta,j)=\beta F_\alpha(N).
\]
Now
\[
  d_{m+1}-d_m=10^{2d_m}+1>(10^{d_m})^2,
\]
so Pantano's Lemma~B.1 implies that $\sum_{j=1}^N\varepsilon(\beta,j)$ tends to infinity
\cite[Appendix~B]{Pantano1995}. Hence $F_\alpha(N)\to\infty$.  For each $r\in\NN$, choose $K$ with $F_\alpha(K)>r$. Because $\alpha$ is irrational, a sufficiently small neighborhood of $\alpha$ contains no rational with denominator at most $K$ whose crossing could change one of
the best lower approximations used in the first $K$ summands. The strict inequality $F_\gamma(K)>r$ therefore persists throughout a rational open interval containing $\alpha$. Thus, $\alpha\in U_r$ (see \eqref{eq:dense-uniform}) for every $r$, and $\alpha\in\mathcal A$.

For the comparison algorithm, the case $a=0$ is immediate.
Assume $a\ge1$. Choose $m$ minimal such that
\[
  Q:=10^{d_m}>10N,
  \qquad
  \frac SQ:=\sum_{i=1}^m10^{-d_i}.
\]
We will show that $a/b<\alpha$ is equivalent to the integer
comparison $aS<bQ$.

The fraction $S/Q$ is reduced, since $S\equiv1\pmod{10}$. As $a\le10N<Q$, equality $aS=bQ$ would imply $Q\mid a$, which is impossible. Therefore
\[
  \left|b-a\frac SQ\right|\ge\frac1Q.
\]
On the other hand, $d_{m+1}>2d_m$ gives
\[
  0<\beta-\frac SQ
  <\sum_{k=d_{m+1}}^\infty10^{-k}
  =\frac{10}{9}10^{-d_{m+1}}
  <\frac1{Q^2}.
\]
Multiplying by $a<Q$, we see that replacing $S/Q$ by $\beta$ changes $b-aS/Q$ by less than $1/Q$, and hence cannot change its sign. Thus
\[
  \frac ab<\alpha
  \quad\Longleftrightarrow\quad
  a\beta<b
  \quad\Longleftrightarrow\quad
  aS<bQ.
\]

Finally, this truncation can be computed in polynomial time. Since $N\ge1$, we have $m\ge2$, and minimality gives $10^{d_{m-1}}\le10N$. Hence
\[
  d_m=d_{m-1}+10^{2d_{m-1}}+1
  \le\log_{10}(10N)+(10N)^2+1
  =O(N^2).
\]
Iterating the recurrence and constructing $S$ and $Q$ therefore takes polynomial time in $N$, with all intermediate integers having $O(N^2)$ bits. The final integer comparison has the same
polynomial bound.
\end{proof}

\section{Finite equivalents for irrational coefficients}
\label{app:colouring-details}
\begin{lemma}[Colouring estimate]
\label[lemma]{lem:colouring-estimates}
Let $\alpha>2$ be irrational, and put
$k:=\lfloor\alpha\rfloor$ and $\ell:=2k+2$.
If $S_\alpha(\ell)<\binom{\ell}{2}$, then, for every
$n\ge\ell$,
\[
  S_\alpha(n)
  <\binom{\ell}{2}
   +\left(\frac{\ell}{2}-\frac1{\ell-1}\right)(n-\ell).
\]
\end{lemma}

\begin{proof}
Suppose that $\alpha>k+2/3$.
The rational approximations with denominators
one, two, and three give
\[
  S_\alpha(\ell)
  \ge k\ell+\frac12+\frac23(\ell-2)
  =\binom{\ell}{2}+\frac{\ell-5}{6}
  >\binom{\ell}{2},
\]
contrary to the hypothesis; here $\ell\ge6$.
Since $\alpha$ is irrational, it follows that
\[
  \alpha<k+\frac23
  <k+1-\frac1{2k+1}
  =\frac{\ell}{2}-\frac1{\ell-1}.
\]
Using $r_\alpha(j)<\alpha$, we obtain, for $n\ge\ell$,
\begin{align*} 
  S_\alpha(n)
  \le S_\alpha(\ell)+\alpha(n-\ell)
  & <\binom{\ell}{2}
   +\left(\frac{\ell}{2}-\frac1{\ell-1}\right)(n-\ell). \qedhere 
\end{align*} 
\end{proof}

\begin{lemma}[Clique cutoffs]
\label[lemma]{lem:clique-cutoffs}
Let $k\ge2$ be an integer and let $\alpha\in(k,k+1)$
be irrational. Put
\[
  \ell:=2k+2,
  \qquad
  \beta_k:=k+\frac12+\frac{1}{4\lfloor k/2\rfloor+2}.
\]
Then
\[
  S_\alpha(\ell)<\binom{\ell}{2}
  \quad\Longleftrightarrow\quad
  \alpha<\beta_k.
\]
\end{lemma}

\begin{proof}
Put $d:=2\lfloor k/2\rfloor+1$.
Then $d\ge3$ is odd, $\ell\in\{2d,2d+2\}$, and
\[
  \beta_k=k+\frac12+\frac1{2d}.
\]

Suppose first that $\alpha>\beta_k$. We have $r_\alpha(1)=k$ and $r_\alpha(j)\ge k+1/2$ for $j\ge2$. Moreover, $\beta_k-k=(d+1)/(2d)$ in its reduced form has denominator $d$, so $r_\alpha(j)\ge\beta_k$ for $j\ge d$. Consequently,
\[
  S_\alpha(\ell)
  \ge k\ell+\frac{\ell-1}{2}+\frac{\ell-d+1}{2d}
  =\binom{\ell}{2}+\frac{\ell-2d+1}{2d}
  >\binom{\ell}{2}.
\]

Now suppose that $\alpha<\beta_k$.
If
\[
  \frac12<\frac ab<\frac12+\frac1{2d},
  \qquad 1\le b\le\ell,
\]
then
\[
  d(2a-b)<b\le2d+2.
\]
Since $2a-b$ is a positive integer with the same parity as $b$,
the only possibilities are
\[
  \begin{array}{lll}
    2a-b=1,& b\ge d+2,
      &\displaystyle \frac ab-\frac12
        \le\frac1{2(d+2)},\\[4pt]
    2a-b=2,& b=2d+2,
      &\displaystyle \frac ab-\frac12
        =\frac1{2(d+1)}.
  \end{array}
\]
Applying these bounds to the rational approximations defining
$r_\alpha(j)-k$ gives
\[
  r_\alpha(j)\le
  \begin{cases}
    k+\dfrac12,
      &2\le j\le d+1,\\[4pt]
    k+\dfrac12+\dfrac1{2(d+1)},
      &d+2\le j\le\ell.
  \end{cases}
\]
The second bound is strict at $j=d+2$, since the case $2a-b=2$ requires denominator $2d+2>d+2$.
Together with $r_\alpha(1)=k$, this yields
\[
  S_\alpha(\ell)
  <k\ell+\frac{\ell-1}{2}+\frac{\ell-d-1}{2(d+1)}
  \le k\ell+\frac{\ell}{2}
  =\binom{\ell}{2},
\]
where the second inequality uses $\ell\le2d+2$.
This completes the calculation.
\end{proof}

\begin{proof}[Proof of \Cref{prop:graph-intervals}]
Put $k:=\lfloor\alpha\rfloor$.
For $\ell\ge4$, put
\[
  D_\ell(n):=\frac{(\ell+1)(\ell-2)n-\ell(\ell-3)}{2(\ell-1)}=\binom{\ell}{2}
  +\left(\frac{\ell}{2}-\frac1{\ell-1}\right)(n-\ell).
\]
An $\ell$-critical graph has chromatic number $\ell$ and every proper subgraph (not necessarily induced) is $(\ell-1)$-colourable. The Kostochka--Yancey bound states that every such graph on $n$ vertices has at least $D_\ell(n)$ edges \cite[Theorem~4]{KYdensity}. Every finite graph that is not $(\ell-1)$-colourable contains an $\ell$-critical subgraph.

We first record a criterion for $(\ell-1)$-colourability
that will be used in both parameter ranges. Every finite subgraph $\struct{H}$ of $\struct{M}_\alpha$ satisfies $e(\struct{H})\le S_\alpha(|\struct{H}|)$. Consequently, if $S_\alpha(n)<D_\ell(n)$ for all $n\ge\ell$, then no finite subgraph is $\ell$-critical, and every finite subgraph is $(\ell-1)$-colourable. Compactness then gives $\struct{M}_\alpha\to\struct{K}_{\ell-1}$.

Suppose first that $1<\alpha<5/3$. Since $r_\alpha(1)=1$, $r_\alpha(2)\le3/2$, and $r_\alpha(j)<5/3$ for all $j$, for $n\ge4$ we have
\[
  S_\alpha(n)
  <1+\frac32+(n-2)\frac53
  <\frac{5n-2}{3}
  =D_4(n).
\]
Applying the above criterion with $\ell=4$, we obtain
$\struct{M}_\alpha\to\struct{K}_3$.
Together with \Cref{thm:graph-hardness}, this gives
$\struct{M}_\alpha\longleftrightarrow\struct{K}_3$.

Now suppose that $\alpha>2$, and put $\ell:=2k+2$. By \Cref{thm:graph-hardness},
$\struct{K}_{\ell-1}\in\mathcal C_\alpha$. All proper induced subgraphs of $\struct{K}_\ell$ therefore belong to $\mathcal C_\alpha$, so
\[
  \struct{K}_\ell\in\mathcal C_\alpha
  \quad\Longleftrightarrow\quad
  S_\alpha(\ell)\ge\binom{\ell}{2}.
\]
If this inequality holds, $\struct{K}_\ell$ embeds into $\struct{M}_\alpha$.
\Cref{thm:graph-hardness} gives the reverse homomorphism. Hence,  $\struct{M}_\alpha\longleftrightarrow\struct{K}_\ell$.

If the inequality fails, the numerical estimate proved in \Cref{lem:colouring-estimates} gives
\[
  S_\alpha(n)
  <\binom{\ell}{2}
   +\left(\frac{\ell}{2}-\frac1{\ell-1}\right)(n-\ell)
  =D_\ell(n)
  \qquad(n\ge\ell).
\]
Applying the above criterion gives
$\struct{M}_\alpha\to\struct{K}_{\ell-1}$.
Together with \Cref{thm:graph-hardness}, this yields
$\struct{M}_\alpha\longleftrightarrow\struct{K}_{\ell-1}$.

By \Cref{lem:clique-cutoffs}, $S_\alpha(\ell)<\binom{\ell}{2}$ holds exactly when $\alpha<\beta_k$, giving the stated cases. In each case, $\struct{M}_\alpha$ is homomorphically equivalent to a finite clique with at least three vertices. Its CSP is therefore the corresponding graph colouring problem and is \NP-complete.
\end{proof}

\end{document}